\documentclass[a4paper,12pt]{article}

\usepackage[utf8]{inputenc}
\usepackage[english]{babel}
\usepackage{amsthm}
\usepackage{bbold}
\usepackage{tikz}
\usetikzlibrary{decorations.pathmorphing,patterns,calc}
\usepackage{amssymb}
\usepackage{tabularx} 
\usepackage{amsmath}  
\usepackage{graphicx} 
\usepackage[margin=1in,letterpaper]{geometry} 
\usepackage[numbers,sort&compress]{natbib} 
\usepackage[final]{hyperref} 
\usepackage{cleveref} 
\usepackage{caption}
\usepackage{subcaption}
\usepackage{csquotes}

\hypersetup{
    colorlinks=true,       
    linkcolor=blue,        
    citecolor=blue,        
    filecolor=magenta,     
    urlcolor=blue          
}

\newcommand{\im}{{\rm i}}
\newcommand{\s}{\mathbb{S}^2}
\newcommand{\td}{\text{d}}
\newcommand{\tD}{\frac{\text{d}^4p}{(2\pi)^4}}
\newcommand{\tDalt}{\frac{\text{d}^3p}{(2\pi)^3}}
\newcommand{\lm}{\lambda^{-}}
\newcommand{\lp}{\lambda^{+}}
\newcommand{\sL}{\sigma_{\Lambda}}
\newcommand{\sk}{\sigma_k}
\newcommand{\intinf}{\int_{-\infty}^{\infty}}
\newcommand{\intinfz}{\int_{0}^{\infty}}

\begin{document}

\title{Quantum Schwarzschild-(A)dS Black Holes: Unitarity and Singularity Resolution}
\author{Steffen Gielen\footnote{\href{mailto:s.c.gielen@sheffield.ac.uk}{s.c.gielen@sheffield.ac.uk}}\; and Sofie Ried\footnote{\href{mailto:sried1@sheffield.ac.uk}{sried1@sheffield.ac.uk}}
\\
\\{\em School of Mathematical and Physical Sciences,} 
\\{\em University of Sheffield,  Hounsfield Road, Sheffield S3 7RH, UK}}
\date{\today}

\maketitle

\begin{abstract}
We consider the canonical quantisation of spherically symmetric spacetimes within unimodular gravity, leaving sign choices in the metric general enough to include both the interior and exterior Schwarzschild--(Anti-)de Sitter spacetime. In unimodular gravity the cosmological constant appears as an integration constant analogous to a total energy, and the quantum Wheeler--DeWitt equation takes the form of a Schr\"odinger equation in unimodular time. We discuss self-adjoint extensions of the Schr\"odinger-like Hamiltonian arising from the requirement of unitarity in unimodular time, and identify a physically motivated one-parameter family of extensions. For semiclassical states we are able to derive analytical expressions for expectation values of the metric, representing a quantum-corrected, nonsingular extension of the classical Schwarzschild--(A)dS geometry which describes a quantum transition between asymptotic black hole and white hole states. The sign of the self-adjoint extension parameter corresponds to the allowed sign of the black hole/white hole mass, and so it can be chosen to ensure that this mass is always positive. We also discuss tunnelling states which allow for a change in the sign of the mass, but which are not semiclassical in high-curvature regions. Our mechanism for singularity resolution and the explicit form of the quantum-corrected metric can be compared to other proposals for black holes in quantum gravity, and in the asymptotically AdS case can be contrasted with holographic arguments.
\end{abstract}

\section{Introduction}

Black holes are among the most fascinating predictions of classical general relativity, and their observational study is a current topic of major interest \citep{abbott_observation_2016,EventHorizonTelescope:2019dse}. They are also expected to serve as windows into what might lie beyond general relativity, whether some modified gravitational theory \citep{Berti:2015itd} or a theory of quantum gravity \citep{Calmet:2014dea}. Their semiclassical description leads us to questions such as what happens at the classical singularity or whether the singularity should be resolved by quantum effects, or what is the final state of Hawking evaporation. Indeed, the black hole information paradox arising from Hawking's original calculations \citep{Hawking:1975vcx} has led to a particularly strong debate into what quantum gravity could do to restore standard principles of quantum mechanics or whether a modification of quantum mechanics is needed \citep{Hawking:1976ra, Susskind:1993if, Almheiri:2012rt, Penington:2019npb}. It is often believed that these issues are connected and explaining or resolving the black hole singularity will help to resolve the question of information loss.

In this article we ask how quantum mechanics on its own can resolve the classical singularity inside a black hole. This is an old question, and many proposals for how to answer it have been made. Given the complications of full quantum gravity, most studies are based on relatively simple models, in particular the spherically symmetric Schwarzschild--(Anti-)de Sitter spacetime. For this model, the interior of the black hole is a spatially homogeneous Kantowski--Sachs solution \citep{kantowskisachs}, which can be quantised as a minisuperspace model; this is also the setup we will work in here. In the exterior, time evolution is replaced by ``radial evolution'' in a spacelike direction. The fate of the classical singularity then depends on how exactly the quantum theory is defined, and what criteria are applied for deciding that a singularity is resolved. Those criteria implicitly use a particular interpretation of quantum mechanics in the context of gravity, and hence are open to debate \citep{singres1,singres2,singres3}. In symmetric cosmological or black hole models, curvature singularities are typically associated to zeros or divergences in metric components, which appear as boundaries of the configuration space of geometries (minisuperspace). This means that in the quantum theory boundary conditions are required, as already discussed by DeWitt \citep{dewitt_quantum_1967}. DeWitt's proposal was to demand that the wavefunction vanishes on such singular boundaries, which was interpreted as a vanishing probability of encountering a singularity, leading to a nonsingular theory. Without a concrete Hilbert space and probability interpretation, it is not clear how such an interpretation is justified (see, e.g., \citep{ashtekarsingh}). In the case of a black hole the singularity is to the future of any observer, hence the boundary condition could be seen as a final rather than initial condition.

Recently, a number of different proposals have been made to connect this issue of boundary conditions to the black hole information paradox \citep{horowitz_black_2004,bouhmadi-lopez_annihilation--nothing_2020,perry_future_2021,perry_no_2022}. In \citep{horowitz_black_2004} Horowitz and Maldacena  suggested a ``totally reflecting" final state of the black hole interior, scattering the state of the infalling Hawking radiation in such a way that the total evolution of collapsing matter into Hawking radiation is unitary (for further discussion of this proposal see, e.g., \citep{gottesman_comment_2004}). In the proposal of \citep{bouhmadi-lopez_annihilation--nothing_2020}, if the wavefunction is assumed to remain bounded there is no freedom to choose a boundary condition at the singularity but a semiclassical state is interpreted as including two opposite arrows of time, leading to ``annihilation'' of singular geometries. Finally, the recent papers \citep{perry_future_2021,perry_no_2022} build on the Horowitz--Maldacena final state proposal using arguments about the wavefunction similar to DeWitt's together with a specific inner product (which is not positive definite) to advocate a probabilistic interpretation indicating that the singularity may be resolved. Again, all these works are based on some assumptions of how quantum mechanics should be interpreted in the context of gravity.

Many interpretational issues of canonical quantum gravity, regarding evolution, observables and possible relation with measurement outcomes, are connected to the problem of time \citep{ishampot,kucharpot,anderson_problem_2012}. The Wheeler--DeWitt equation, derived from the Hamiltonian constraint of general relativity, does not contain a Schr\"odinger-like background time. One can take the view, as we will do here, that dynamical statements can only be formulated in relational terms, i.e., as the evolution of observables with respect to a particular ``clock''. Then statements about probabilities of certain outcomes are also, in general, relational and depend on the clock. This relational viewpoint was recently studied by Hartnoll \citep{hartnoll_wheeler-dewitt_2022} in the context of different clock choices for semiclassical quantum states describing planar black holes. It was found that for these clocks the semiclassical approximation holds all the way to the singularity, hence there is no strong departure from classical evolution and no singularity resolution. The analysis was extended to charged planar black holes in \cite{Blacker:2023ezy}.

The formalism we will use in the following is similar to Hartnoll's, but our conclusions will be different because we use a different type of clock: as discussed by Gotay and Demaret  \citep{gotay_quantum_1983,gotay_remarks_1997}, whether or not a singularity is resolved by quantum effects depends on whether the theory is expected to be unitary with respect to a ``slow'' clock or with respect to a ``fast'' clock. Slow clocks reach the singularity in finite time, and it is here that unitarity (the requirement of well-defined evolution for all times) clashes with the idea that a solution should terminate in a singularity. More specifically, as we will see, unitarity in such a clock leads to a new proposal for a (non-unique) boundary condition at the singularity, rather different from the previous ones based on probability arguments. This proposal is then explicitly clock-dependent \citep{gotay_quantum_1983,gotay_remarks_1997,gielen_singularity_2020,gielen_unitarity_2022,gielen_unitarity_2022-1}, and so in a way we have moved the question of finding boundary conditions to that of finding the clock variable with respect to which we require unitarity. 

We will be working in unimodular gravity, a theory that is classically equivalent to general relativity but promotes the cosmological constant to a global degree of freedom arising as an integration constant from the trace-free Einstein equations. Many formulations of unimodular gravity exist \citep{AndersonFinkelstein,Buchmuller,unruh_unimodular_1989,henneaux_cosmological_1989,carballo-rubio_unimodular_2022} but they tend to reduce to the same theory in minisuperspace. The key feature in our context is that these theories introduce a specific notion of preferred time \citep{unruh_time_1989}, which we will utilise to define our clock. It is then with respect to this clock that we require unitarity. The interior of a planar black hole has recently been studied imposing unitarity with respect to unimodular time and it was found that the singularity is resolved \citep{gielen_black_2024}. In this paper we will not restrict ourselves to the interior of the black hole and our conclusions will be somewhat different, since we will study the full Schwarzschild--(Anti-)de Sitter solution rather than the simpler planar version. In our case, we will show that applying our procedure only to the interior of the black hole, or equivalently to Kantowski--Sachs cosmology, creates an additional boundary condition at the horizon, which makes the theory harder to analyse, which is why we will include the exterior in our work.

After quantising Schwarzschild-(Anti-)de Sitter spacetime using the Wheeler--DeWitt approach and imposing unitary unimodular time evolution, we arrive at a theory in which all classical singularities have been replaced by nonsingular quantum evolution past the classical singularity. Hence, requiring unitarity in a suitable clock can serve as a mechanism for resolving singularities, as already seen in other contexts previously \citep{gotay_quantum_1983,gotay_remarks_1997,gielen_singularity_2020,gielen_unitarity_2022,gielen_unitarity_2022-1, gielen_black_2024}. This mechanism can be compared with ideas for resolving black hole singularities in other approaches to quantum gravity such as asymptotic safety \cite{Falls:2010he,Held:2019xde}, loop quantum gravity \cite{Gambini:2013hna,Kelly:2020uwj,Ashtekar2023}, or spin foam models \cite{Soltani:2021zmv} as well as more phenomenological models of singularity-free black holes \cite{bardeen,Dymnikova:1992ux,Hayward:2005gi}. In the context of asymptotically Anti-de Sitter black holes, our findings can be contrasted with arguments using holography which state that black hole singularities should {\em not} be resolved; see in particular the conjecture of \citep{engelhardt_holographic_2016}. The arguments supporting this conjecture rely on the fact that the quantum theory can be dually described by a boundary conformal field theory (CFT). In our approach, no such dual description exists since the quantum theory contains information not present in the asymptotic limit, in particular the non-unique choice of boundary condition at the singularity. Our requirement of unitarity in unimodular time also has no obvious analogue in holography. Hence, these seem to be fundamentally different pictures of what quantum gravity says about singularities. We will return to this point in the conclusions.

Our second key result is that the theory permits the existence of semiclassical states consisting exclusively of either positive-mass black/white holes or negative-mass black/white holes, where the same state describes a spacetime transitioning from black to white hole of the same mass, and the sign of the mass depends on the choice of self-adjoint extension or boundary condition at the singularity. Hence, the non-uniqueness of the boundary condition translates into qualitatively different definitions of the quantum theory, even in the semiclassical limit. If the self-adjoint extension is chosen so that only positive mass solutions exist, this resolves the issue that would otherwise occur when both positive and negative mass black hole states coexist in a theory without singularities. In the absence of singularities, the cosmic censorship conjecture does not preclude the existence of negative-mass black holes, and it seems that such objects could be formed from the vacuum which would be unstable \citep{HorowitzMyers}. A new explanation would be required for why such objects have not been observed. Going away from the semiclassical theory, we also find states describing the tunnelling from a black hole spacetime to a white hole, with the mass switching sign; one should take this result with a grain of salt, as a more complete theory of quantum gravity might change the calculation substantially in the highly quantum regime. 

This paper is structured into five sections. After this introduction we will begin by introducing the classical theory in \cref{sec:clth}, presenting our minisuperspace variables and briefly reviewing unimodular gravity. In \cref{sec:cqt} we will quantise the theory, impose unitarity and study the possible self-adjoint extensions of the Hamiltonian. Here we will see that some self-adjoint extensions correspond to a theory with positive mass semiclassical solutions and others correspond to a theory with negative mass semiclassical solutions. Afterwards we will calculate expectation values and their variance in \cref{sec:aqt} to confirm the semi-classical nature of the state and compare it to a tunnelling state. In \cref{sec:dis} we will summarise our results and discuss our findings. In this paper we will work in natural units with $\hbar = c = 8\pi G = 1$, unless explicitly said otherwise.

\section{Classical theory}
\label{sec:clth}
We will be working in a minisuperspace approach, reducing degrees of freedom of our metric by making symmetry assumptions prior to quantising. In the interior of the Schwarzschild--(Anti-)de Sitter black hole the spacetime is homogeneous and spherically symmetric, but not isotropic, and we can write our line element as
\begin{align}
\label{equ:lineel}
    \td s^2 = -\frac{\eta(t)N(t)^2}{\xi(t)}\td t^2 + \frac{\xi(t)}{\eta(t)} \td z^2 + \eta(t)^2\td\Omega^2\,.
\end{align}
Here $\td\Omega^2$ is the standard metric on $\s$ and the coordinate $z$ parametrises the real line. The $g_{tt}$ and $g_{zz}$ components of the metric are chosen in this way to simplify future equations. We  see that if $\xi$ changes sign the metric is still Lorentzian but the role of the timelike coordinate switches from $t$ to $z$, corresponding to crossing a horizon in Schwarzschild--(Anti-)de Sitter solutions that involve regions of positive and negative $\xi$. In contrast, as we will see below, $\eta\rightarrow 0$  corresponds to a curvature singularity where 2-spheres shrink to a point, and we will always assume that $\eta\ge 0$. $g_{tt}$ cannot change sign independently of $g_{zz}$ since $N(t)^2\ge 0$. This metric parametrisation is similar to what was done for isotropic cosmology in \citep{halliwell_derivation_1988}, except that a sign change of $\xi$ here does not change the sign convention of the metric. 

Another way of interpreting this form of metric is to view it as a homogeneous but anisotropic cosmology. From this point of view we are studying Kantwoski--Sachs cosmology if $N^2$, $\xi$ and $\eta$ are strictly positive. Quantisation of Kantowski--Sachs cosmology has been studied for example in \citep{uglum_quantum_1992,conradi_quantum_1995}, and  \citep{uglum_quantum_1992} uses the same parametrisation in terms of $\eta$ and $\xi$.  Both papers solve the Wheeler--DeWitt equation for Kantowski--Sachs cosmology in vacuum or, e.g., including pressureless dust. They discuss several choices of boundary condition such as the Hartle--Hawking ``no boundary" condition  \citep{hartle_wave_1983}, the Vilenkin ``outgoing flux'' proposal \citep{vilenkin_quantum_1988} and the symmetric initial condition \citep{conradi_quantum_1991,conradi_initial_1992}.   We will recover the classical and some of the quantum solutions discussed in these papers but we will interpret them differently.

We will now calculate the Hamiltonian constraint for this form of metric in the context of unimodular gravity, and analyse the classical solutions. Expressing the system in the Hamilton--Jacobi formalism connects closely to the process of canonical quantisation discussed in \cref{sec:cqt}.

\subsection{Hamiltonian formulation}
To derive the Hamiltonian, we begin by considering the action. There are several ways of defining a unimodular version of general relativity; here we will work with the Henneaux--Teitelboim version \citep{henneaux_cosmological_1989} which includes an extra field that can be interpreted as time. This is in contrast with other formulations \citep{AndersonFinkelstein,Buchmuller} that use a fixed volume form which breaks the diffeomorphism invariance of the action to a subgroup of transverse diffeomorphisms, the ones that do not alter the determinant of the metric. In unimodular gravity {\em \`a la} Henneaux--Teitelboim general diffeomorphism invariance is preserved, the action has the form
\begin{align}
    S_{HT}[g_{\mu\nu},\Lambda,T^{\mu}] = \frac{1}{2}\int\td^4x[\sqrt{-g}R-2\Lambda(\sqrt{-g}-\partial_{\mu}T^{\mu})]\,.
\end{align}
Here $R$ denotes the Ricci scalar and $g$ is the determinant of the metric. The cosmological constant is promoted to a field $\Lambda(x^{\mu})$. The new variables $T^{\mu}$ act as Lagrange multipliers ensuring that $\Lambda$ is a constant. The vector density $T^{\mu}$ is only well defined up to a divergenceless vector density, since $T^\mu\rightarrow T^\mu+J^\mu$ with $\partial_\mu J^\mu=0$ is a local gauge symmetry of the action. Thus $T^\mu$ only has one degree of freedom that is not gauge. The first two terms give the Einstein--Hilbert action with $\Lambda$ appearing as a field. Variation with respect to $\Lambda$ gives the following equation of motion, also called the unimodular condition:
\begin{align}
    \sqrt{-g} = \partial_{\mu}T^{\mu}\,.
\end{align}
This constraint on the metric determinant is the analogue of fixing a volume form in other approaches to unimodular gravity, with the difference is that $T^{\mu}$ is a dynamical field. The introduction of this new degree of freedom avoids fixing the metric determinant by hand. 

A particular feature of this theory is that it includes a preferred notion of time which we call {\em unimodular time}. Let $\{\Sigma(t)\}_t$ be a foliation of spacetime into spacelike hypersurfaces and let $n_{\mu}$ be the future-directed covector normal to the hypersurface, unimodular time can then be defined as 
\begin{align}
    T_{\Lambda}(t) := \int_{\Sigma(t)}\td^3x\; n_{\mu}T^{\mu}\,.
\end{align}
The change in unimodular time has a very nice interpretation as the spacetime volume between two hypersurfaces. If $U$ is a submanifold of spacetime with boundary $\partial U = \Sigma(t_i)\sqcup\Sigma(t_f)$ we can define the spacetime volume of $U$ as follows:
\begin{align}
    \mathcal{V}(U) := \int_U\td^4x\sqrt{-g} &= \int_{U}\td^4x\,\partial_{\mu}T^{\mu}\\
    &= \int_{\Sigma(t_f)}\td^3x\,n_{\mu}T^{\mu} - \int_{\Sigma(t_i)}\td^3x\,n_{\mu}T^{\mu}\\
    &= T_{\Lambda}(t_f)-T_{\Lambda}(t_i)\,.
\end{align}
In the first line we have used the unimodular condition. It should be noted that the notion of unimodular time fully characterises time only when accompanied by a fixed foliation as, in general,
two distinct hypersurfaces may agree on the ``time'' $T_\Lambda(t)$ assigned to them \citep{kuchar_does_1991} (contributions from the four-volume between these hypersurfaces can come with positive and negative signs, and add up to zero). A foliation effectively selects one representative from each equivalence class of hypersurfaces.
Since we are working in minisuperspace we have already chosen a specific foliation. A nice introduction to Henneaux--Teitelboim theory and the notion of unimodular time can be found in the MSc thesis \citep{etkin_quest_2023}. This ``objective'' notion of time in unimodular gravity motivates us to use $T_{\Lambda}$ as our clock.

Using our minisuperspace metric (\ref{equ:lineel}) the action simplifies. We also assume that the fields $T^\mu$ and $\Lambda$ are only function of $t$, so that spatial derivative terms $\partial_i T^i$ vanish. After adding suitable (Gibbons--Hawking--York) boundary terms to remove the dependence on second time derivatives, the action then becomes
\begin{align}
    S_{HT}[N,\xi,\eta,\Dot{\xi},\Dot{\eta},\Lambda,\Dot{T}] = V\int\td t\left[N-\frac{1}{N}\Dot{\xi}\Dot{\eta}-N\Lambda\eta^2+\Lambda\Dot{T}\right]
\end{align}
where we write $T$ instead of $T^0$. Since the minisuperspace variables do not depend on the spatial coordinates, we have performed the integrals over all coordinates but $t$, and the 3-dimensional coordinate volume is denoted by $V$. This assumes that the $z$ dimension has been compactified in some way, e.g., either by restricting the range of $z$ to some finite interval or by assuming it is periodic. We can absorb $V$ into a variable redefinition 
\begin{align}
\label{equ:defetaxi}
   VN\rightarrow N,\quad VT\rightarrow T, \quad V^2\xi \rightarrow \xi
\end{align}
so that now $T = T_{\Lambda}$ denotes the unimodular time as defined above, and find an action
\begin{align}
    S_{HT} = \int\td t \left[N-\frac{1}{N}\Dot{\xi}\Dot{\eta}-N\Lambda\eta^2+\Lambda\Dot{T}\right]\,.
\end{align}
We can introduce conjugate momenta
\begin{align}
\label{equ:conjmomenta}
    \pi_{\eta} = -\frac{1}{N}\Dot{\xi}\,, \quad \pi_{\xi} = -\frac{1}{N}\Dot{\eta}\,, \quad \pi_T = \Lambda\,.
\end{align}
As stated above the cosmological constant is conjugate to $T$. Recall from (\ref{equ:lineel}) that depending on the sign of $\xi$, the coordinate $t$ can be either timelike or spacelike. In the spacelike case, the notion of conjugate momenta introduced here is a generalisation of the usual notion, since we treat quantities like $\dot\xi$ as velocities even though they correspond to derivatives in a radial direction. This generalisation of Hamiltonian methods to  spacelike ``evolution'' is discussed, e.g., in \cite{deBoer:2000cz} in the context of AdS/CFT, where the generalised momenta are interpreted as variations of the action. While the physical interpretation of the resulting quantities may be subtle, the mathematical framework for spacelike evolution is always well-defined in our effectively one-dimensional symmetry-reduced setup.

The Hamiltonian is given by
\begin{align}
\label{equ:ham}
    H = -N\left[\pi_{\eta}\pi_{\xi}+1-\eta^2\Lambda\right]\,.
\end{align}
The action does not depend on the time derivative of the lapse meaning its conjugate momentum $\pi_N$ is constrained to vanish. For consistency, its time derivative $\Dot{\pi}_N$, given by the derivative of the Hamiltonian with respect to the lapse, also has to vanish:
\begin{align}
\label{equ:hamcon}
    0 = \Dot{\pi}_N = - \frac{\partial H}{\partial N} = \pi_{\eta}\pi_{\xi}+1-\eta^2\Lambda =: \mathcal{C}\,.
\end{align}
We obtain the Hamiltonian constraint $\mathcal{C}$. The Hamiltonian can be written as $H=-N\mathcal{C}$ and therefore also has to vanish. The Hamiltonian is now the starting point for calculating the classical solutions as well as quantising the minisuperspace theory.

\subsection{Classical solutions}
In order to calculate the classical solutions we will initially take the lapse $N$ to be a positive constant $N_0$. Then the Lagrangian equations of motion are
\begin{align}
\label{equ:hamcongauge}
    0 &= \frac{1}{N_0^2}\Dot{\xi}\Dot{\eta}+1-\Lambda\eta^2\,,\\
\label{equ:eometa}
    \Ddot{\eta} &= 0\,,\\
    \label{equ:eomxi}
    \Ddot{\xi} &= 2N_0^2\Lambda\eta\,,\\
    \label{equ:eomT}
    \Dot{T} &= N_0\eta^2\,,\\
    \Dot{\Lambda} &= 0\,.
\end{align}
The last equation states that $\Lambda$ has to be a constant and we will treat it as such in the following. The first one is simply the Hamiltonian constraint. There are two classes of classical solutions depending on whether $\Dot{\eta}$ vanishes or not. 

Let us start with the case $\Dot{\eta} \neq 0$. In that case \cref{equ:eometa} dictates that 
\begin{align}
    \eta(t) = \eta_0+N_0k t\,,\quad k\neq 0\,.
\end{align}
The solution for $\xi$ can then be found by integrating the Hamiltonian constraint \eqref{equ:hamcongauge}:
\begin{align}
    \xi(t) = \frac{\Lambda k}{3}N_0^3t^3+\Lambda\eta_0N_0^2t^2+\left(\frac{\eta_0^2}{k}\Lambda-\frac{1}{k}\right)N_0t+\xi_0\,.
\end{align}
A simple integration of \cref{equ:eomT} gives the expression for the unimodular time,
\begin{align}
    T(t) = \frac{k^2}{3}N_0^3t^3+\eta_0kN_0^2t^2+\eta_0^2N_0t+T_0\,.
\end{align}
Since $T^{\mu}$ is only defined up to a divergenceless vector we can add any constant to it and set $T_0 = 0$. By shifting $t$ we can set $\eta_0 = 0$. The solutions then become
\begin{align}
\label{equ:clsol}
    \eta(t) = kN_0t,\quad \xi(t) = \frac{\Lambda}{3}kN_0^3t^3-\frac{1}{k}N_0t+\xi_0,\quad T(t) = \frac{k^2}{3}N_0^3t^3\,.
\end{align}
By our previous assumption that $\eta\ge 0$, the solution is only defined for $t\ge 0$ if $k> 0$, and only defined for $t\le 0$ if $k<0$. Hence, depending on the sign of $k$ the curvature singularity at $t=0$ is either in the future or past, if we take $t$ to indicate the arrow of time. While $t$ is a coordinate time, \cref{equ:clsol} shows that its sign matches that of $T$, our unimodular time used to define a global time. We will take $T$ to define the arrow of time and hence the solution describes a white hole for $k>0$ and a black hole for $k<0$.
The metric also becomes singular for $\xi = 0$, however we will show that this does not correspond to a curvature singularity, but is a relic of our choice of foliation. The vanishing of $\xi$ indicates a horizon, either an event horizon or a cosmological horizon. 

\begin{figure}
\centering
\begin{tikzpicture}
\colorlet{mydarkpurple}{blue!40!red!50!black}
\colorlet{mylightpurple}{mydarkpurple!80!red!6}

\tikzset{declare function={%
  kruskal(\x,\c)  = {\fpeval{asin( \c*sin(2*\x) )*2/pi}};
}}

 \draw[blue!40!red!80!black,line width=0.4,samples=20,smooth,variable=\x,domain=-8:8] 
        plot(\x,{-4*kruskal(\x*pi/16,0.5)})
        plot(\x,{ 4*kruskal(\x*pi/16,0.5)});

\draw[blue!40!red!80!black,line width=0.4,samples=20,smooth,variable=\y,domain=-4:4] 
        plot({-4*kruskal(\y*pi/16,0.5)},\y)
        plot({ 4*kruskal(\y*pi/16,0.5)},\y);

\draw[blue!40!red!80!black,line width=0.4,samples=20,smooth,variable=\y,domain=-4:0] 
        plot({4*kruskal(\y*pi/16,0.5)+8},\y);

\draw[blue!40!red!80!black,line width=0.4,samples=20,smooth,variable=\y,domain=0:4] 
        plot({4*kruskal(\y*pi/16,0.5)-8},\y);

\draw[blue!40!red!80!black,line width=0.4,samples=20,smooth,variable=\y,domain=0:4] 
        plot({-4*kruskal(\y*pi/16,0.5)+8},\y);

\draw[blue!40!red!80!black,line width=0.4,samples=20,smooth,variable=\y,domain=-4:0] 
        plot({-4*kruskal(\y*pi/16,0.5)-8},\y);

\node[blue!40!red!80!black] at (4,1.6) {$z_{max}$};
\node[blue!40!red!80!black] at (4,-1.6) {$z_{min}$};

\fill[fill=mylightpurple, opacity=0.7]
  plot[smooth,domain=-8:8,variable=\x] 
    (\x,{-4*kruskal(\x*pi/16,0.5)})
  --
  plot[smooth,domain=-8:8,variable=\x] 
    (\x,{ 4*kruskal(\x*pi/16,0.5)})
  -- cycle;

\coordinate (IIIr) at (4/3,4) ;
\coordinate (IIIl) at (-4/3,4) ;
\coordinate (IVr) at (4/3,-4) ;
\coordinate (IVl) at (-4/3,-4) ;
\coordinate (Vll) at (-8,4) ;
\coordinate (Vlr) at (-28/3,4) ;
\coordinate (VIll) at (-8,-4) ;
\coordinate (VIlr) at (-28/3,-4) ;
\coordinate (Vrl) at (28/3,4) ;
\coordinate (Vrr) at (8,4) ;
\coordinate (VIrl) at (28/3,-4) ;
\coordinate (VIrr) at (8,-4) ;
\coordinate (left) at (-8,0);
\coordinate (right) at (8,0);

\fill[fill=mylightpurple, opacity=0.7]
  plot[smooth,domain=-4:4,variable=\y] 
    ({4*kruskal(\y*pi/16,0.5)},\y)
  -- (IIIr) -- (IIIl) --
  plot[smooth,domain=-4:4,variable=\y] 
    ({-4*kruskal(\y*pi/16,0.5)},\y)
-- (IVr) -- (IVl)
  -- cycle;

\fill[fill=mylightpurple, opacity=0.7]
  plot[smooth,domain=-4:0,variable=\y] 
    ({4*kruskal(\y*pi/16,0.5)+8},\y)
  -- (right) -- (VIrr) -- (VIrl)
  -- cycle;

\fill[fill=mylightpurple, opacity=0.7]
  plot[smooth,domain=0:4,variable=\y] 
    ({4*kruskal(\y*pi/16,0.5)-8},\y)
  -- (Vlr) -- (Vll) -- (left)
  -- cycle;

\fill[fill=mylightpurple, opacity=0.7]
  plot[smooth,domain=0:4,variable=\y] 
    ({-4*kruskal(\y*pi/16,0.5)+8},\y)
  -- (Vrl) -- (Vrr) -- (right)
  -- cycle;

\fill[fill=mylightpurple, opacity=0.7]
  plot[smooth,domain=-4:0,variable=\y] 
    ({-4*kruskal(\y*pi/16,0.5)-8},\y)
  -- (left) -- (VIll) -- (VIlr)
  -- cycle;

\node (I)    at ( 4,0)   {I : $\xi < 0$};
\node (II)   at (-4,0)   {II: $\xi < 0$};
\node (III)  at (0, 2.5) {III: $\xi > 0$};
\node (IV)   at (0,-2.5) {IV: $\xi > 0$};
\node (V) at (-7,2.5) {$\xi > 0$};
\node (VI) at (-7,3) {V:};
\node (VII) at (7,2.5) {$\xi > 0$};
\node (VIII) at (7,3) {V:};
\node (IX) at (-7,-2.5) {VI:};
\node (X) at (-7,-3) {$\xi> 0$};
\node (XI) at (7,-2.5) {VI:};
\node (XII) at (7,-3) {$\xi > 0$};

\path  
  (II) +(90:4)  coordinate (IItop)
       +(-90:4) coordinate (IIbot)
       +(0:4)   coordinate (IIright)
       +(180:4) coordinate (IIleft)
       ;
\draw (IIleft) -- 
          node[midway, below, sloped] {$\xi = 0$}
      (IItop) --
          node[midway, below, sloped] {$\xi=0$}
      (IIright) -- 
          node[midway, above, sloped] {$\xi=0$}
      (IIbot) --
          node[midway, above, sloped] {$\xi = 0$}
      (IIleft) -- cycle;

\path 
   (I) +(90:4)  coordinate (Itop)
       +(-90:4) coordinate (Ibot)
       +(180:4) coordinate (Ileft)
       +(0:4)   coordinate (Iright)
       ;
\draw  (Ileft) -- 
          node[midway, below, sloped] {$\xi = 0$}
      (Itop) --
          node[midway, below, sloped] {$\xi=0$}
      (Iright) -- 
          node[midway, above, sloped] {$\xi=0$}
      (Ibot) --
          node[midway, above, sloped] {$\xi = 0$}
      (Ileft) -- cycle;

\draw[decorate,decoration=zigzag] (IItop) -- (Itop)
      node[midway, above, inner sep=2mm] {$\eta=0$};

\draw[decorate,decoration=zigzag] (IIbot) -- (Ibot)
      node[midway, below, inner sep=2mm] {$\eta=0$};

\path
    (Ibot) +(0:4) coordinate (botright)
    (Itop) +(0:4) coordinate (topright)
    (IIbot) +(180:4) coordinate (botleft)
    (IItop) +(180:4) coordinate (topleft);

\draw   (Ibot) -- node[midway, below]    {$\eta = \infty$}
    (botright) -- (Iright) -- (topright) -- node[midway, above]    {$\eta = \infty$}
    (Itop);

\draw   (IIbot) -- node[midway, below]    {$\eta = \infty$}
    (botleft) -- (IIleft) -- (topleft) -- node[midway, above]    {$\eta = \infty$}
    (IItop);

\end{tikzpicture}
\caption{Conformal diagram of Schwarzschild--de Sitter spacetime. The sign of $\xi$ is indicated in each region. $\eta$ is positive everywhere and vanishes at the singularities. The left and right boundary are identified. The chosen coordinates do not cover all of the spacetime, in particular they are restricted to the shaded regions indicating $z\in [z_{min},z_{max}]$, as we have compactified this dimension. For $k>0$ parts of the regions IV, I and V with $z_{min}\leq z\leq z_{max}$ are covered, for $k<0$ parts of the regions VI, I and III with $z_{min}\leq z\leq z_{max}$.}
\label{fig:condiag}
\end{figure}
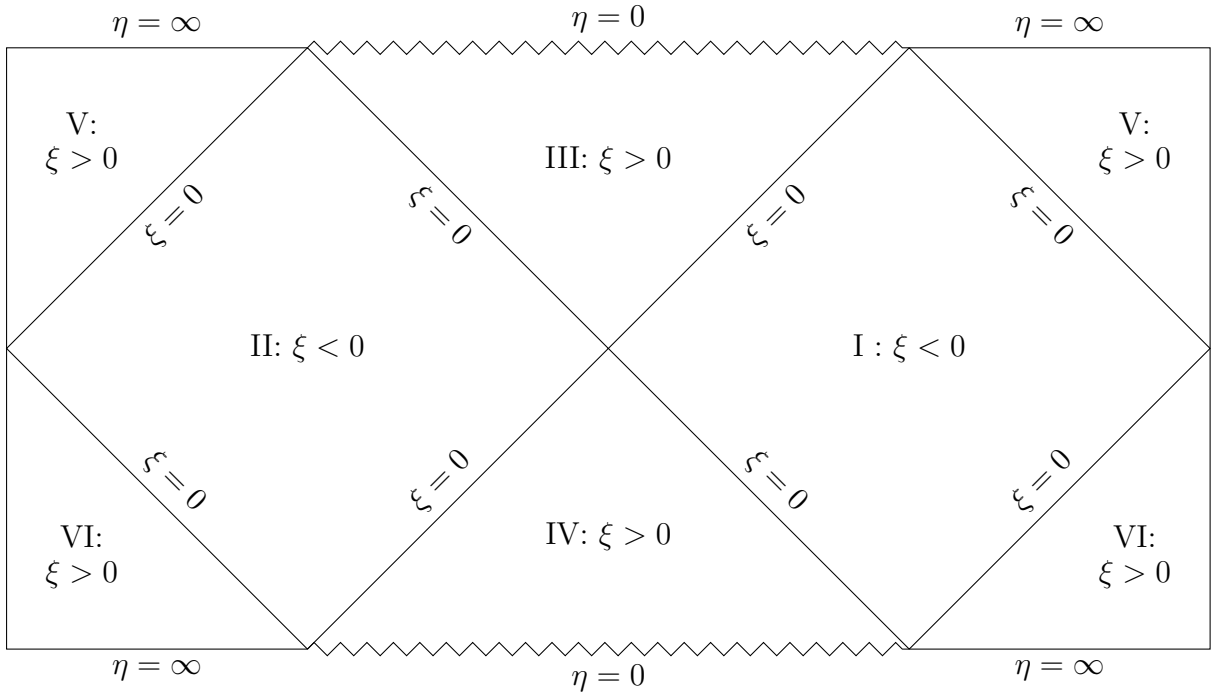

This spacetime structure is illustrated in figure \ref{fig:condiag} for the positive $\Lambda$ case, i.e., for the Schwarzschild--de Sitter solution.
There are analytical continuations of the spacetime to the left and the right of the picture, so we identify the left and right boundary to avoid infinite successions of black and white holes. The regions of negative $\xi$ correspond to the exterior of the black and white hole inside of cosmological horizons. The regions with $\xi > 0$ are either inside an event horizon or outside of the cosmological horizon. The squiggly line at the top is the singularity of the black hole, the one at the bottom is the singularity of the white hole. It should be noted that our choice of coordinates does not cover all of Schwarzschild--de Sitter spacetime, as it is not defined at the horizons. In the case $k<0$, our solution describes a black hole spacetime and covers the region VI outside of the cosmological horizon, exterior region I and the black hole interior III. Further discussion of the conformal diagram can be found in \cite{PhysRevD.15.2738}. For the cases $\Lambda\le 0$, the global structure of the spacetime is simpler since there are no cosmological horizons. The global structure of the Schwarzschild--Anti-de Sitter spacetime is discussed in \cite{maldacena_cool_2013,Socolovsky:2017nff}, for example. In these cases future/past spacelike infinity ($\eta=\infty$) is replaced by timelike infinity ($\Lambda<0$) or null infinity ($\Lambda=0$). 

The solutions in \cref{equ:clsol} depend on three parameters $k$, $\xi_0$ and $N_0$, but $N_0$ is arbitrary and we can use the remaining gauge freedom of rescaling the coordinates $t$ and $z$ to understand the significance of $k$ and $\xi_0$. A constant rescaling of the time coordinate $t\rightarrow\alpha t$ corresponds to a rescaling of the lapse $N_0\rightarrow N_0/\alpha$, and a rescaling $z\rightarrow \beta z$ corresponds to rescaling $\xi\rightarrow \xi/\beta$ and $N_0\rightarrow\sqrt{\beta}N_0$ in \cref{equ:lineel}. $N_0$ can always be chosen at will. Starting with the expression for $\xi(t)$, we can now try to eliminate the dependence on $k$:
\begin{align}
    \xi(t) = \frac{\Lambda}{3}kN_0^3t^3-\frac{1}{k}N_0t+\xi_0 &\stackrel{N_0 = \frac{1}{|k|}}{\longrightarrow} \xi(t) = \frac{\Lambda \,\text{sgn}(k)}{3k^2}t^3-\frac{\text{sgn}(k)}{k^2}t+\xi_0\\
    \label{equ:k2xi0}
    &\stackrel{\cdot k^2}{\longrightarrow} \xi(t) = \frac{\Lambda}{3}\text{sgn}(k)\,t^3-\text{sgn}(k)\,t+k^2\xi_0\,.
\end{align}
The rescaling performed in the second step is not  obviously allowed in our minisuperspace model, but as it corresponds to a rescaling of a spatial coordinate in the metric, it maps between minisuperspace solutions that describe the same metric and hence the same solution of the full theory. After applying these transformations, we have $\eta(t)=\text{sgn}(k)\,t$ and given that the metric only includes the ratio $\xi/\eta$, it will no longer depend separately on $k$ and $\xi_0$ but only on the combination $k^2\,\text{sgn}(k)\,\xi_0$. The dependence on $\text{sgn}(k)$ can further be eliminated by sending $t\rightarrow -t$ in the case $k<0$ to ensure that the $t$ coordinate always takes positive values. As discussed above this amounts to mapping a black hole solution to the corresponding white hole solution by reversing the arrow of time. We then have $\eta(t)=t$, and the line element is given by
\begin{align}
    \td s^2 = -\left[\frac{\Lambda}{3}t^2-1+\frac{k^2\xi_0}{t}\right]^{-1}\td t^2 + \left[\frac{\Lambda}{3}t^2-1+\frac{k^2\xi_0}{t}\right]\td z^2+t^2\td \Omega^2\,,
\end{align}
which is the usual Schwarzschild--(Anti-)de Sitter line element inside the horizon. One can see that the parameters $k$ and $\xi_0$ are related to the black/white hole mass $M$ via $k^2\xi_0 = M/4\pi$, and they do not have independent physical significance. On the other hand, we see that the innocuous integration constant $\xi_0$, whose presence is already expected from the fact that the Hamiltonian does not depend explicitly on $\xi$, has physical significance: solutions with different values of $\xi_0$, related by a simple constant translation in minisuperspace, are interpreted as having different mass.

\begin{figure}
     \centering
     \begin{subfigure}[b]{0.48\textwidth}
         \centering
         \includegraphics[width=\textwidth]{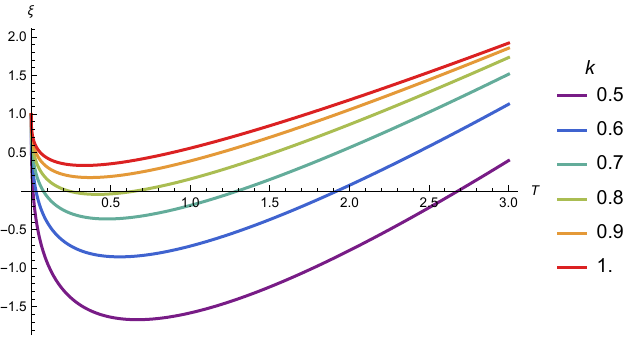}
         \caption{$\xi(T)$ for $\Lambda = 1$ and\\$0.5\leq k\leq 1$.}
     \end{subfigure}
     \hfill
     \begin{subfigure}[b]{0.48\textwidth}
         \centering
         \includegraphics[width=\textwidth]{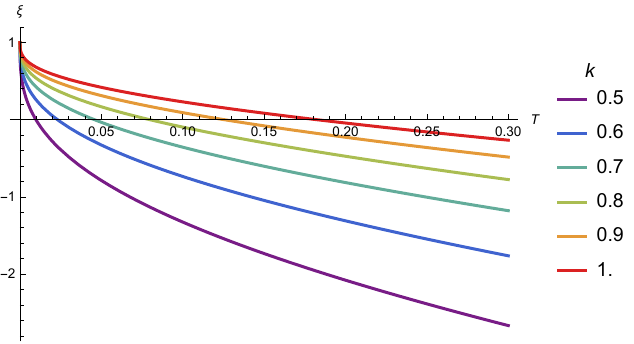}
         \caption{$\xi(T)$ for $\Lambda = -1$ and\\$0.5\leq k\leq 1$.}
     \end{subfigure}
        \caption{Visualisation of $\xi(T)$ for different choices of $\Lambda$ and $k$ and positive white hole mass $M=4\pi k^2 \xi_0$, with $\xi_0=1$. For positive $\Lambda$ there are solutions with two horizons as well as cosmological solutions. For negative $\Lambda$ all solution have an event horizon, the location of which depends on the mass.}
        \label{fig:clsol1}
\end{figure}

So far we have expressed our classical solutions as functions of a particular time coordinate $t$, but these solutions $\xi(t)$, $\eta(t)$ and $T(t)$ are gauge-dependent and not well-defined observables in our quantum theory. To construct observables we have to consider relational quantities such as $\xi(T = T^*)$, which tells us the value of $\xi$ conditioned on a unimodular time $T=T^*$. This is what we mean when saying we will use unimodular time $T$ as our clock. Since the relation between $T$ and $t$ is easily invertible we can write down the classical expressions $\eta(T)$ and $\xi(T)$:
\begin{align}
\label{equ:relcl}
    \eta(T) = \sqrt[3]{3kT}\,,\quad \xi(T) = \frac{\Lambda}{k}T - \sqrt[3]{\frac{3T}{k^5}}+\xi_0\,.
\end{align}
These expressions are visualised for several choices of $k$ and $\Lambda$ in \cref{fig:clsol1} for $\xi_0=1$. The behaviour of $\xi(T)$ depends on the sign of the cosmological constant $\Lambda$. For $\Lambda > 0$ there are solutions that stay positive for all times (to be interpreted as anisotropic cosmologies) as well as solutions that have two horizons, an event horizon and a cosmological horizon. The condition for a solution to have horizons and therefore describe a black or white hole spacetime is
\begin{align}
    k^2\xi_0 < \frac{2}{3\sqrt{\Lambda}}\,.
\end{align}
In the case of equality $k^2\xi_0=\frac{2}{3\sqrt{\Lambda}}$, the cosmological and event horizon coincide and $\xi=0$ at exactly one time $T$. On the other hand if $\Lambda < 0$, then all solutions have an event horizon. This behaviour of the Schwarzschild--(Anti-)de Sitter metric is well-known. Since the clock variable $T$ is simply a power of coordinate time $t$ the qualitative behaviour of the solutions is independent of whether one considers them a function of coordinate or unimodular time. Notably $\eta(0) = 0$ for all solutions and $\xi(0) = \xi_0$. Going back to the definition of $\xi$, \cref{equ:lineel}, this means that $g_{zz}$ has to diverge at $T=0$. In the case of a spacetime with horizons, this behaviour describes the singularity at the centre of the black/white hole. To show that this is not a feature of the coordinates chosen one can calculate curvature invariants such as the Kretschmann scalar $K = R^{\mu\nu\rho\sigma}R_{\mu\nu\rho\sigma}$, where $R_{\mu\nu\rho\sigma}$ is the Riemann tensor, and show that these diverge as well. Using the formula given in \citep{gkigkitzis_kretschmann_2014}, the Kretschmann scalar for the Schwarzschild--(Anti-)de Sitter metric can be calculated to be
\begin{align}
\label{eq:kretschmann}
    K = \frac{12k^4\xi_0^2}{\eta^6}+\frac{8\Lambda^2}{3} = \frac{3M^2}{4\pi^2\eta^6}+\frac{8\Lambda^2}{3}\,.
\end{align}
We have used the relation $k^2\xi_0 = M/4\pi$ to the black/white hole mass to write the scalar in a more familiar way. Clearly this diverges at $T=0$ and nowhere else.

The second class of solutions is characterised by $\Dot{\eta} = 0$. Looking at \cref{equ:conjmomenta} one sees that this implies $\pi_{\xi} = 0$. In this case \cref{equ:hamcongauge} gives
\begin{align}
    \eta = \frac{1}{\sqrt{\Lambda}}\,.
\end{align}
Such solutions only exist for $\Lambda > 0$. The expressions for $\xi$ and $T$ are  then given by integrating \cref{equ:eomxi} and \cref{equ:eomT} respectively:
\begin{align}
    \xi(t) &= \sqrt{\Lambda}N_0^2t^2+v_{\xi}N_0t+\xi_0\,,\\
    T(t) &= \frac{N_0}{\Lambda}t+T_0\,.
\end{align}
Using the shift symmetry of coordinate time $t$, the rescaling symmetry of the coordinates $z$ and $t$, the gauge freedom in unimodular time $T$ as well as the residual gauge freedom of the lapse $N_0$, we can fix all integration constants except $\Lambda$. There are then only three physically distinct solutions:
\begin{align}
    \eta(t) = \frac{1}{\sqrt{\Lambda}}\,,\quad T(t)=\frac{1}{\Lambda}t\,,\quad\xi(t) = \sqrt{\Lambda}t^2-\epsilon\quad\text{with}\quad\epsilon\in\{-1,0,1\}\,.
\end{align}
By rescaling $t$ and $z$ we can rewrite the resulting metric in the following way:
\begin{align}
    \td s^2 = \frac{1}{\Lambda}\left[-(\epsilon-t^2)\td z^2+\frac{\td t^2}{\epsilon-t^2}+\td\Omega^2\right]\,.
\end{align}
This metric describes Nariai spacetime \cite{nariai1950some,nariai1951new}, which can be described as the product spacetime $dS_2\times S^2 $ of two-dimensional de Sitter space with a 2-sphere of constant radius. For $\epsilon = 1$ the de Sitter part describes the static patch of de Sitter space, $\epsilon = 0$ describes the flat slicing of de Sitter and $\epsilon = -1$ the closed slicing. For $\epsilon \in \{0,1\}$  the usual part of de Sitter space covered by these coordinates is given either for $t\in(0,\infty)$ or $t\in(-\infty,0)$, whereas for the closed slicing $t$ covers the whole real line. As before the $z$ dimension is compactified and can therefore only take values in some interval $I$.

For this class of solutions the Kretschmann scalar is $K = 4\Lambda^2$. Notably there are no singularities in these solutions. In the quantum theory these solutions will not be normalisable, since $\pi_{\xi} = 0$ translates to the fact that the wavefunction has to independent of $\xi$ (see below).


Another way to arrive at the classical solutions is the Hamilton--Jacobi formalism,  a reformulation of classical mechanics that describes the motion of a system in terms of a generating function. The Hamilton function is given by
\begin{align}
    H(T,\eta,\xi,\pi_T,\pi_{\eta},\pi_{\xi}) = \pi_{\eta}\pi_{\xi}+(1-\pi_T\eta^2) = 0\,.
\end{align}
A function $S$ that fulfils the property
\begin{align}
    H\left(T,\eta,\xi,\frac{\partial S}{\partial T},\frac{\partial S}{\partial\eta},\frac{\partial S}{\partial \xi}\right) = 0
\end{align}
is called a generating function. In our case one is given by \citep{uglum_quantum_1992}
\begin{align}
    S = k\xi+\frac{1}{k}\left(\frac{\Lambda}{3}\eta^3-\eta\right)+\Lambda T\,,
\end{align}
as can be verified by a short computation. The equations of motion in the Hamilton--Jacobi formalism can then be computed to be
\begin{align}
    \pi_{\xi} = \frac{\partial S}{\partial \xi} = k, \quad \pi_{\eta} = \frac{\partial S}{\partial \eta} = \frac{1}{k}(\Lambda\eta^2-1), \quad \pi_T = \frac{\partial S}{\partial T} = \Lambda\,,\\
    Q_k = \frac{\partial S}{\partial k} = \xi - \frac{1}{k^2}\left(\frac{\Lambda}{3}\eta^3-\eta\right),\quad Q_{\Lambda} = \frac{\partial S}{\partial \Lambda} = \frac{1}{3k}\eta^3-T\,.
    \label{eq:HamiltonJacobi}
\end{align}
In the Hamilton--Jacobi formalism all the information about the dynamics is in the generating function $S$. The resulting equations of motion are trivial, in the sense that $Q_k$ and $Q_{\Lambda}$ are constants. The constant $Q_{\Lambda}$ constitutes a shift in $T$ which is a gauge transformation, therefore we can set $Q_{\Lambda}=0$ without loss of generality. However, $Q_k$ is important for the interpretation of the resulting solutions: by setting $Q_{\Lambda}=0$ and $Q_k = \xi_0$ we arrive at the same expressions as in \cref{equ:relcl}, with the black or white hole mass again related to the combination $k^2\xi_0$.  Again, notice that the integration constant $\xi_0$ acquires physical significance through this interpretation. Solutions with the same momenta $k$ and $\Lambda$ but different $\xi_0$ have the same generating function $S$ and are hence indistinguishable in the Hamilton--Jacobi formalism.

The advantage of the Hamilton--Jacobi formalism is that it is easy to connect to the quantum setting through the generating function and through the fact that the dynamical equations are already expressed in relational form, without dependence on an arbitrary time coordinate $t$ or gauge parameter $N$. Note that the $\pi_{\xi} = 0$ solutions are not included here as they correspond to $k = 0$ for which the generating function is not well-defined. This will not bother us, as these solutions are not normalisable in the quantum theory. 

\section{Constructing the quantum theory}
\label{sec:cqt}
Our understanding of the classical solutions, developed in the previous section, will now help us to construct a quantum theory for our minisuperspace model. The starting point for the quantisation is the Hamiltonian given in \cref{equ:ham}. We will rewrite it slightly:
\begin{align}
    H = -N\left[\pi_{\eta}\pi_{\xi}+1-\eta^2\Lambda\right] = -\Tilde{N}\left[\frac{1}{\eta^2}\pi_{\eta}\pi_{\xi}+\frac{1}{\eta^2}-\Lambda\right] = -\Tilde{N}\Tilde{\mathcal{C}}\,.
\end{align}
We have redefined the lapse function and the constraint $\mathcal{C}$ in order to isolate $\Lambda$; since the lapse is arbitrary this does not change the classical dynamics. We can now write the new constraint in terms of a minisuperspace metric for the $(\eta,\xi)$ subspace as
\begin{align}
    \tilde{\mathcal{C}} &= -(H_S+\Lambda) := g^{\mu\nu}\pi_{\mu}\pi_{\nu}+\frac{1}{\eta^2}-\Lambda\,,\\
    g^{\mu\nu} &= \frac{1}{2\eta^2}\begin{pmatrix}
0 & 1 \\
1 & 0 
\end{pmatrix}\,,\quad \pi_1 = \pi_{\eta}\,,\quad \pi_2 = \pi_{\xi}\,.
\label{eq:mssmetric}
\end{align}
When quantising, we need to represent the constraint $H_S+\Lambda=0$ as a requirement for physical states to be annihilated by the operator versions of $H_S$ and $\Lambda$. We will think of states as wavefunctions $\psi(\eta,\xi,T)$. In general the issue of operator ordering then introduces some ambiguity. Here we will adopt the Laplace--Beltrami ordering and map $g^{\mu\nu}\pi_{\mu}\pi_{\nu}$ to the Laplace--Beltrami operator on the $(\eta,\xi)$ part of minisuperspace:
\begin{align}
    g^{\mu\nu}\pi_{\mu}\pi_{\nu}\quad \mapsto\quad -\hbar^2 \Delta = -\frac{\hbar^2}{\sqrt{-g}}\partial_{\mu}\left(g^{\mu\nu}\sqrt{-g}\partial_{\nu}\right) = -\frac{\hbar^2}{\eta^2}\partial_{\eta}\partial_{\xi}\,.
\end{align}
This operator ordering is uniquely determined by demanding invariance under coordinate transformations in the $(\eta,\xi)$ coordinates, apart from a possible additive term involving the Ricci scalar of the minisuperspace metric \citep{halliwell_derivation_1988}. Since the metric (\ref{eq:mssmetric}) is flat this additive term vanishes, and the operator ordering is unique under the assumption of covariance in the $(\eta,\xi)$ subspace. The quantised form of the constraint, known as the Wheeler--DeWitt equation, takes on the structure of a Schrödinger equation\footnote{The resulting Schr\"odinger-like quantum theory can be interpreted in the framework of the Page--Wootters formalism \cite{pagewootters,wootters} as describing the evolution of a ``system'' (the geometry parametrised by $\eta$ and $\xi$) with respect to a ``clock'' $T$. This is a fully covariant treatment which agrees with other quantisation methods such as Dirac quantisation \cite{hoehn_trinity_2021}.}:
\begin{align}
\label{equ:WdW}
    \im\hbar\partial_T\psi = \hat{H}_S\psi\quad \text{with}\quad\hat{H}_S = \frac{\hbar^2}{\eta^2}\partial_{\eta}\partial_{\xi}-\frac{1}{\eta^2}\,.
\end{align}
Here, the factors of $\hbar$ are explicitly included to emphasise the analogy with the Schrödinger equation. For simplicity, $\hbar$ will again be set to unity in subsequent discussions. This structure arises because the cosmological constant $\Lambda$, conjugate to the unimodular time $T$, appears linearly in the constraint. By imposing the constraint $-\Lambda=H_S$ at the quantum level, we see that the observable value of $\Lambda$ corresponds to minus the expectation value of the Schrödinger Hamiltonian $\hat{H}_S$. Consequently, we require that $\hat{H}_S$ is an observable and therefore self-adjoint within our framework. This requirement applies regardless of whether $T$ is timelike or spacelike. In the specific case of a black/white hole spacetime with positive mass, $T$ is timelike near the singularity. As we will see later, imposing unitarity in $T$ can be interpreted as enforcing a boundary condition at the singularity ($\eta = 0$). Thus, in such scenarios, the requirement for self-adjointness of $\hat{H}_S$ has an additional motivation from the requirement of ensuring unitary time evolution.

Since we know a generating function for the classical Hamilton--Jacobi equation it is straightforward to find quantum solutions. An eigenbasis of the Schrödinger-type operator $\hat{H}_S$ can be indexed by the eigenvalues $\Lambda$ and the degeneracy parameter $k$ as
\begin{align}
    \psi_{\Lambda,k}(\eta,\xi,T) = e^{\im S[\Lambda,k]} = e^{\im\left[k\xi+\frac{1}{k}\left(\frac{\Lambda}{3}\eta^3-\eta\right)+\Lambda T\right]}\,.
\end{align}
These solutions can be seen as semiclassical in the sense that their phase is given by the generating function $S[\Lambda,k]$ and the amplitude varies slowly (indeed is even constant). However they are not approximate but exact solutions to the Wheeler--DeWitt equation, in the sense of solutions to a differential equation. Boundary conditions will be discussed shortly. By our comments below \cref{eq:HamiltonJacobi}, it should be clear that even for fixed $k$ and $\Lambda$ these solutions do {\em not} have a unique classical interpretation in terms of a Schwarzschild--(Anti-)de Sitter solution with given mass, since the integration constant $\xi_0$ is undetermined.

A general solution of \cref{equ:WdW} can be decomposed in this basis as
\begin{align}
\label{equ:generalwf}
    \psi( \eta,\xi,T) = \int_{-\infty}^{\infty}\frac{\td\Lambda}{2\pi}\int_{-\infty}^{\infty}\frac{\td k}{2\pi}\,\alpha(\Lambda,k)\,e^{\im\Lambda T}e^{\im\left[k\xi+\frac{1}{k}\left(\frac{\Lambda}{3}\eta^3-\eta\right)\right]}\,.
\end{align}
These solutions exclude the $\pi_{\xi}=0$ case discussed in the previous section, which would be given by
\begin{align}
\label{equ:etacon}
    \psi(\eta,T) = \intinf\frac{\td k}{2\pi}\intinf\frac{\td \Lambda}{2\pi}\,\alpha(\Lambda)\,e^{\im \Lambda T}e^{-\im\left(\eta-\frac{1}{\sqrt{\Lambda}}\right)k}\,.
\end{align}
Such a wavefunction does not depend on $\xi$ and thus fulfils the property $\pi_{\xi} = 0$; it also satisfies \cref{equ:WdW} as can be easily verified. We will however see shortly that such states can not be normalised and are thus not included in the subsequent analysis.

The next step is to restrict the space of solutions of the constraint to a subset on which the Schrödinger-type Hamiltonian $\hat{H}_S$ is self-adjoint and thus the evolution in $T$ is unitary. In order to define self-adjointness we first need to define an inner product. It is natural to choose a Schrödinger-type scalar product,
\begin{align}
\label{equ:ipxxi}
    \langle\psi_1,\psi_2\rangle_T = \int_0^{\infty}\td\eta\int_x^{\infty}\td\xi\,\sqrt{-g}\,\bar\psi_1(\eta,\xi,T)\psi_2(\eta,\xi,T)\,,
\end{align}
where $g = -4\eta^4$ is the determinant of the minisuperspace metric. An interesting question is how to choose the range of $\xi$, which is why we have kept the lower limit $x$ of $\xi$ arbitrary in the definition above. If we want to quantise Kantowski--Sachs spacetime we need to take $\xi\in(0,\infty)$. However, from the classical solutions we know that most solutions have at least one horizon. Therefore imposing unitarity will give us two boundary conditions at the singularity and at the horizon, which are difficult to impose simultaneously. An alternative option, which will turn out to be easier to implement, is to include the exterior region of black/white hole solutions and allow $\xi$ to take arbitrary real values, as we already did when discussing general classical solutions.

Regardless of the choice of the lower bound for $\xi$, the states defined by \cref{equ:etacon} cannot be normalised as they are independent of $\xi$ and delta distributions in $\eta$. Hence these states will not be part of our quantum theory.

We now need to find a domain $\mathcal{D}(\hat{H}_S)\subset\mathcal{H}$ on which the Schr\"odinger-type Hamiltonian $\hat{H}_S$ is self-adjoint, where $\mathcal{H}$ is the usual quantum-mechanical Hilbert space of square-integrable wavefunctions (at some given time $T$) with respect to (\ref{equ:ipxxi}).  For arbitrary $\psi,\phi\in\mathcal{H}$ we would find 
\begin{align}
\label{equ:boundaryterms}
\begin{split}
    \langle \phi,\hat{H}_S\psi\rangle - \langle \hat{H}_S\phi,\psi\rangle = &\intinfz\td\eta\left((\partial_{\eta}\Bar{\phi}(\eta,x))\psi(\eta,x)-\Bar{\phi}(\eta,x)\partial_{\eta}\psi(\eta,x)\right)\\
    &+ \int_x^{\infty}\td\xi\left((\partial_{\xi}\Bar{\phi}(0,\xi))\psi(0,\xi)-\Bar{\phi}(0,\xi)\partial_{\xi}\psi(0,\xi)\right)\,.
\end{split}
\end{align}

To make the boundary terms vanish, we can take $\mathcal{D}(\hat{H}_S)$ to be the subspace of square-integrable functions that are twice weakly differentiable and fulfil the property $\psi(0,\xi)=0$ and $\psi(\eta,x)=0$. Note that the second property is automatically fulfilled if $x = -\infty$. Then $\hat{H}_S$ is symmetric on its domain, i.e.,
\begin{align}
    \langle \hat{H}_S\phi,\psi\rangle = \langle \phi,\hat{H}_S\psi\rangle\quad\forall\,\phi,\,\psi\in\mathcal{D}(\hat{H}_S)\,.
\end{align}
However, $\hat{H}_S$ defined in this way is not self-adjoint\footnote{A basic introduction into the important distinction between symmetric and self-adjoint operators in quantum mechanics can be found in \cite{Bonneau:1999zq}.}. The adjoint $\hat{H}_S^*$, while formally corresponding to the same differential operator $\hat{H}_S^* = \frac{\hbar^2}{\eta^2}\partial_{\eta}\partial_{\xi}-\frac{1}{\eta^2}$, is defined on the larger domain
\begin{align}
    \mathcal{D}(\hat{H}^*_S) := \{\phi_1\in\mathcal{H}|\,\exists \,\phi_2\in\mathcal{H},\,\text{s.t.}\,\langle\hat{H}_S\psi,\phi_1\rangle = \langle\psi,\phi_2\rangle\,\forall\psi\in\mathcal{D}(\hat{H}_S)\}\,,
\end{align}
which in our case is dense in the Hilbert space $\mathcal{H}$. We thus have the relation $\hat{H}_S\subset\hat{H}_S^*$, as the domain of the adjoint is strictly larger than the domain of $\hat{H}_S$. We now want to find extensions of the Hamiltonian that are self-adjoint, i.e., operators $\hat{A}$ such that
\begin{align}
    \hat{H}_S \subset \hat{A} = \hat{A}^* \subseteq \hat{H}_S^*\,.
\end{align}
In other words, we need to relax the boundary condition in $\mathcal{D}(\hat{H}_S)$ to find a larger space of allowed wavefunctions. In general there will be more than one self-adjoint extension and the theories described by different extensions could behave very differently.

\subsection{Restricting to black/white hole interiors}
\label{sec:interior}
This case is very similar to the one discussed in \cite{Daughton:1998aa}, where it is demonstrated that self-adjoint extensions of the operator exist based on von Neumann's theorem, which provides a non-constructive proof. Here we explicitly compute a class of self-adjoint extensions.

We will start by defining our Hamiltonian operator on the domain
\begin{align}
    \mathcal{D}_{(\mu,\nu)}(\hat{H}_S) = \left\{\psi\in\mathcal{H}\,|\,\psi(0,\xi)=\psi\left(0,\mu/\xi\right),\,\psi(\eta,0)=\psi\left(\nu/\eta,0\right)\right\}\,.
    \label{eq:newdomain}
\end{align}
We then change coordinates in \cref{equ:boundaryterms} to $u = \log(\eta)$ and $v = \log(\xi)$, and the limits of the integrals become $\pm\infty$. If $\phi$ and $\psi$ are then functions with the symmetries $\psi(0,v)=\psi(0,\mu-v)$ and $\psi(u,0)=\psi(\nu-u,0)$, their derivative will be odd with respect to these transformations, and the integrals vanish from symmetry. Translating this symmetry back to the $(\eta,\xi)$ coordinates we arrive at the domain in \cref{eq:newdomain}, and we see that $\hat{H}_S$ is symmetric on this domain. In order to prove that it is self-adjoint we have to calculate the domain of its adjoint and show that it is equal to its domain defined above. From \cref{equ:boundaryterms}, we have
\begin{align}
\begin{split}
    \langle\phi,\hat{H}_S\psi\rangle = \langle\hat{H}_S\phi,\psi\rangle&+\intinfz\td\xi\,\left((\partial_{\xi}\Bar{\phi}(0,\xi))\psi(0,\xi)-\Bar{\phi}(0,\xi)\partial_{\xi}\psi(0,\xi)\right)\\
    &+\intinfz\td\eta\,\left((\partial_{\eta}\Bar{\phi}(\eta,0))\psi(\eta,0)-\Bar{\phi}(\eta,0)\partial_{\eta}\psi(\eta,0)\right)\,.
\end{split}
\end{align}
The domain of the adjoint is given by all $\phi\in\mathcal{H}$ for which the two integrals vanish for all $\psi\in\mathcal{D}_{(\mu,\nu)}(\hat{H}_S)$. We can again set $u = \log(\eta)$ and $v = \log(\xi)$, and define the functions $\Tilde{\phi}(v) = \Bar{\phi}(0,e^v)$ and $\Tilde{\psi}(v) = \psi(0,e^v)$. Focusing  on the $\xi$ integral first, we are then looking for $\tilde\phi$ such that
\begin{align}
     \intinf\td v\,[\Tilde{\phi}(v)\partial_v\Tilde{\psi}(v)-\partial_v\Tilde{\phi}(v)\Tilde{\psi}(v)] = 0\qquad\forall\,\psi\in \mathcal{D}_{(\mu,\nu)}(\hat{H}_S)\,.
\end{align}
Now let us write $\tilde{\phi}$ and $\tilde{\psi}$ in terms of the Fourier transforms
\begin{align}
    \tilde{\phi}(v) = \intinf \frac{\td k}{2\pi}\,f(k) e^{\im k\left(v-\frac{\mu}{2}\right)}\,,\quad    \tilde{\psi}(v) = \intinf \frac{\td k}{2\pi}\,g(k) e^{\im k\left(v-\frac{\mu}{2}\right)}\,.
\end{align}
By defining the Fourier transformation in this way we ensure that $g(k) = g(-k)$, as this is the symmetry in Fourier space corresponding to the symmetry around $\mu/2$ of $\tilde{\psi}$. We thus have the requirement
\begin{align}
    0 &= \intinf\td v\intinf\frac{\td q}{2\pi}\intinf\frac{\td k}{2\pi}\, \left[\im k f(q)g(k)e^{\im(k+q)\left(v-\frac{\mu}{2}\right)}-\im q f(q)g(k)e^{\im(k+q)\left(v-\frac{\mu}{2}\right)}\right]\,,\\
    &= 2\im\intinf\frac{\td k}{2\pi}\,kf(-k)g(k) = -2\im\intinf\frac{\td k}{2\pi}\,kf_{{\rm odd}}(k)g(k)\,.
    \label{eq:requirement}
\end{align}
The last step uses the fact that $kg(k)$ is odd and therefore, if we split $f=f_{{\rm even}}+f_{{\rm odd}}$, the contribution coming from $f_{{\rm even}}$ vanishes. Eq. (\ref{eq:requirement}) has to hold for all $g(k)$ that are even, in particular we can take $g(k) = k\Bar{f}_{{\rm odd}}(k)$, then we have the condition
\begin{align}
    0 = 2\im\intinf\frac{\td k}{2\pi}|g(k)|^2\,\quad\Rightarrow g(k) \equiv 0 \Rightarrow f_{{\rm odd}}(k) \equiv 0\,.
\end{align}
Thus $\phi$ has to have the symmetry $\phi(0,\xi) = \phi(0,\mu/\xi)$. An analogous calculation for the $\eta$
integral concludes the proof showing that $\phi\in\mathcal{D}_{(\mu,\nu)}(\hat{H}_S)$ and therefore the self-adjointness of the operator on this domain.

We have identified specific self-adjoint extensions of the operator when restricted to the interior of the black/white hole, i.e., to positive $\xi$. However, this class of extensions does not necessarily include all possible self-adjoint extensions. To compute expectation values in a quantum theory defined by a given self-adjoint extension, we need to determine how the associated boundary conditions constrain the functions $\alpha(\Lambda,k)$ in the wavefunction expansion (\ref{equ:generalwf}). As we were not able to explicitly translate the boundary conditions into constraints on $\alpha(\Lambda,k)$, further analysis of this theory needs to be postponed to future work.

\subsection{Including the exterior region}
We saw that restricting $\xi\in\mathbb{R}^+$ is theoretically consistent with maintaining unitary time evolution; however, the practical challenges of implementing this restriction make it unfeasible for detailed analysis. It is also worth highlighting the conceptual difference between the two approaches to the quantisation of a black or white hole spacetime. Restricting $\xi$ to be positive amounts to placing a reflecting boundary at the horizon ensuring that the quantum modes do not exit into the exterior. This would assume that the horizon exhibits quantum effects (as in \cite{gielen_black_2024}). Generically the spacetime curvature at the horizon can be quite low for a large black/white hole, so that one could argue that one still expects general relativity to be a valid description and  quantum effects to be negligible. This reasoning would suggest that we should allow $\xi$ to become negative, so that we do not have any boundary at the horizon. Of course we do not know what happens at the horizon of a black or white hole and both approaches represent in principle different possibilities of what could happen in quantum gravity. Nonetheless, choosing $x= -\infty$ precludes the application of our method to quantising the cosmological Kantowski-Sachs spacetime.

We will thus take the lower limit of the range of $\xi$ to be $x = -\infty$, and find that we can construct all self-adjoint extensions, parametrised by a free function $\chi(k)$. We will first compute the self-adjoint extensions of the Hamiltonian $\hat{H}_S$ by a direct calculation. Then we will use the method of deficiency indices to arrive at the same conclusion, with a little bit more insight about how the operator fails to be self-adjoint prior to imposing boundary conditions.

Imposing unitarity with respect to $T$ means the scalar product $\langle\psi_1,\psi_2\rangle$ should be time-independent. For general wavefunctions (\ref{equ:generalwf}), we can calculate the scalar product to be
\begin{align}
    \langle\psi_1,\psi_2\rangle_T = \int_0^{\infty}\td\eta\int_{-\infty}^{\infty}\td\xi\int \tD\, &2\eta^2e^{\im T(\Lambda_2-\Lambda_1)}\Bar{\alpha}_1(\Lambda_1,k_1)\alpha_2(\Lambda_2,k_2)\nonumber\\
    &\times e^{\im\left[(k_2-k_1)\xi+\left(\frac{\Lambda_2}{k_2}-\frac{\Lambda_1}{k_1}\right)\frac{\eta^3}{3}-\left(\frac{1}{k_2}-\frac{1}{k_1}\right)\eta\right]}\,.
\end{align}
Here and in the following $\int \tD$ stands for the integration over $\Lambda_{1/2}$ and $k_{1/2}$. We can perform the integrals over $\xi$ and  either $k_1$ or $k_2$:
\begin{align}
    \langle\psi_1,\psi_2\rangle_T =&    \int_0^{\infty}\td\eta\int\tDalt\,2\eta^2e^{\im T(\Lambda_2-\Lambda_1)}\Bar{\alpha}_1(\Lambda_1,k)\alpha_2(\Lambda_2,k)e^{\im(\Lambda_2-\Lambda_1)\frac{\eta^3}{3k}}\\
    \begin{split}
    =& \int_0^{\infty}\td\eta\int\tDalt\,\eta^2e^{\im T(\Lambda_2-\Lambda_1)}\Bar{\alpha}_1(\Lambda_1,k)\alpha_2(\Lambda_2,k)e^{\im(\Lambda_2-\Lambda_1)\frac{\eta^3}{3k}}\\
    &+\int_{-\infty}^{0}\td\eta\int\tDalt\,\eta^2e^{\im T(\Lambda_2-\Lambda_1)}\Bar{\alpha}_1(\Lambda_1,-k)\alpha_2(\Lambda_2,-k)e^{\im(\Lambda_2-\Lambda_1)\frac{\eta^3}{3k}}
    \end{split}\\
    \begin{split}
    =&  \int_{-\infty}^{\infty}\td\eta\int\tDalt\,\eta^2e^{\im T(\Lambda_2-\Lambda_1)}e^{\im(\Lambda_2-\Lambda_1)\frac{\eta^3}{3k}}[\Bar{\alpha}_1(\Lambda_1,-k)\alpha_2(\Lambda_2,-k)\\
    &+\Theta(\eta)(\Bar{\alpha}_1(\Lambda_1,k)\alpha_2(\Lambda_2,k)-\Bar{\alpha}_1(\Lambda_1,-k)\alpha_2(\Lambda_2,-k))]\,.
    \end{split}
    \label{equ:dpdef}
\end{align}
The expression $\int\tDalt$ denotes the integration over $\Lambda_{1/2}$ and $k = k_1 = k_2$. In the last expression we see that the first term will result in a delta distribution $\delta(\Lambda_2-\Lambda_1)$ after integration over $\eta$, whereas the second term will result in a Fourier transform of the Heaviside function $\Theta$ and therefore not eliminate the time dependence. Thus we have the boundary condition
\begin{equation}
\Bar{\alpha}_1(\Lambda_1,k)\alpha_2(\Lambda_2,k)-\Bar{\alpha}_1(\Lambda_1,-k)\alpha_2(\Lambda_2,-k) = 0\quad\;
    \Rightarrow \alpha(\Lambda,-k) = e^{\im\chi(k)}\alpha(\Lambda,k)
    \label{equ:alphacon}
\end{equation}
where $\chi(k)$ is an arbitrary odd function of $k$. Different choices for $\chi(k)$ correspond to different self-adjoint extensions of the Hamiltonian. Using a substitution $y = \frac{\eta^3}{3k}$ we can now finish calculating the scalar product under the assumption that the boundary condition is satisfied:
\begin{align}
    \langle\psi_1,\psi_2\rangle_T &= \int_{-\infty}^{\infty}\td y\int\tDalt|k|\Bar{\alpha}_1(\Lambda_1,k)\alpha_2(\Lambda_2,k)e^{\im(\Lambda_2-\Lambda_1)T}e^{\im(\Lambda_2-\Lambda_1)y}\\
    &= \int_{-\infty}^{\infty}\frac{\td\Lambda}{2\pi}\int_{-\infty}^{\infty}\frac{\td k}{2\pi}|k|\Bar{\alpha}_1(\Lambda,k)\alpha_2(\Lambda,k)\,.
\end{align}
This shows that the boundary condition indeed defines a self-adjoint extension of the Hamiltonian. One can also immediately see that the norm of a state satisfying the boundary condition is strictly positive:
\begin{align}
    |\psi|^2 =  \int_{-\infty}^{\infty}\frac{\td\Lambda}{2\pi}\int_{-\infty}^{\infty}\frac{\td k}{2\pi}|k||\alpha(\Lambda,k)|^2 > 0\,.
    \label{eq:normexpr}
\end{align}

As mentioned above we can also use the method of deficiency indices to calculate the self-adjoint extensions. The deficiency indices $n_{\pm}$ of a closed, symmetric operator $\hat{A}$ are defined to be the dimension of the space of eigenfunctions of the adjoint of $\hat{A}$ to the eigenvalues $\pm \im$:
\begin{align}
    n_{+} &:= \text{dim}\,\text{ker}\,(\hat{A}^*-\im\mathbb{1})\,,\\
    n_{-} &:= \text{dim}\,\text{ker}\,(\hat{A}^*+\im\mathbb{1})\,.
\end{align}
Depending on the values of the deficiency indices there are three options \citep{schmudgen_unbounded_2012}:
\begin{itemize}
    \item $n_{-} = n_{+} = 0$: in that case the operator is already self-adjoint.
    \item $n_{-} = n_{+} \neq 0$: in that case the operator is not self-adjoint, but there are self-adjoint extensions. Self-adjoint extensions are in one-to-one correspondence with unitary maps from $\text{ker}\,(\hat{A}^*-\im\mathbb{1})$ to $\text{ker}\,(\hat{A}^*+\im\mathbb{1})$.
    \item $n_{-} \neq n_{+}$: in that case there are no self-adjoint extensions.
\end{itemize}
In our case we know that a general eigenfunction to the eigenvalue $\pm \im$ is given by
\begin{align}
    \phi_{\pm}(\eta,\xi) = \intinf\frac{\td k}{2\pi}\gamma(k)e^{\im\left[k\xi+\frac{1}{k}\left(\mp\frac{\im}{3}\eta^3-\eta\right)\right]}\,.
\end{align}
However, we need to make sure these functions actually belong to the Hilbert space defined by the scalar product (\ref{equ:ipxxi}). A short calculation shows
\begin{align}
    \langle\phi_{\pm},\phi_{\pm}\rangle = \intinfz\frac{\td k}{2\pi}\,k|\gamma(k)|^2\intinfz\td x\,e^{\pm x}-\int_{-\infty}^0\frac{\td k}{2\pi}\,k|\gamma(k)|^2\int_{-\infty}^0\td x\,e^{\pm x}\,.
\end{align}
Depending on the sign of the eigenvalue, either the first or the second term blows up. Consequently for $\phi_{\pm}$ to be in the Hilbert space one needs to require $\gamma(k) \equiv 0\,\forall \pm k>0$. Let $\mathcal{H}_{\gamma}$ be the space of functions $\gamma(k)$ such that
\begin{align}
    \intinf\frac{\td k}{2\pi}|k||\gamma(k)|^2 <\infty\,.
\end{align}
Then we have now seen:
\begin{align}
    \text{ker}\,(\hat{H}^*_S\mp i\mathbb{1}) = \{\phi_{\pm}|\,\gamma\in\mathcal{H}_{\gamma},\,\gamma(k) \equiv 0\,\forall \pm k>0\}\,.
\end{align}
Unitary maps between the eigenspaces are then given by maps in $\mathcal{H}_{\gamma}$ that map functions of positive $k$ to functions of negative $k$ without changing the absolute value, i.e.,
\begin{align}
    \gamma_-(k)\mapsto e^{\im\chi(k)}\gamma_+(-k)\,,\quad k\geq 0\,,
\end{align}
where $\chi(k)$ can be any real function.
We have seen in the direct calculation that this relation gives exactly the self-adjoint extensions we defined in \cref{equ:alphacon}. The method using deficiency indices comes to the same conclusion (thankfully) and gives a little insight into why the self-adjoint extensions have the form they do, and how the operator fails to be self-adjoint in the first place. Pictorially speaking the boundary conditions given by unitary maps between $\text{ker}\,(\hat{H}^*_S-i\mathbb{1})$ and $\text{ker}\,(\hat{H}^*_S+i\mathbb{1})$ ensure that the contributions of eigenfunctions to imaginary eigenvalues $\lambda$ are cancelled out by contributions of the eigenfunctions to the conjugate eigenvalue $\Bar{\lambda}$. Interestingly negative/positive $k$ correspond to modes with eigenvalues with negative/positive imaginary part. In the classical theory $k>0$ solutions corresponded to white hole spacetimes and $k<0$ solutions to black hole spacetimes. Our analysis here shows that we need to combine both to ensure unitarity in the quantum theory.

There is an undetermined odd function $\chi(k)$ that corresponds to different choices of self-adjoint extension. Every extension defines a different quantum theory of our minisuperspace model. While there is no reason to expect different extensions to give the same results, we can use symmetry arguments to restrict ourselves to more specific, physically interesting quantum theories. Going back to the Wheeler--DeWitt equation (\ref{equ:WdW}), one can see that it is invariant under the following transformation:
\begin{align}
\label{equ:symham}
    \eta \rightarrow -\eta,\quad \xi \rightarrow \beta-\xi\,,\quad \beta\in\mathbb{R}\,.
\end{align}
This is similar to a parity transformation in quantum mechanics, but with an additional shift in $\xi$. In quantum mechanics it is very common to require wavefunctions to be invariant under transformations such as parity reversal if the Hamiltonian has the same symmetry \cite{Bonneau:1999zq}. We will take the analogous step and require our wavefunctions to be invariant under the transformation described in \cref{equ:symham}. The wavefunction cannot be invariant under this transformation for arbitrary $\beta$, as that would imply independence from $\xi$. However, for a fixed $\beta$, specifically $\beta = 2\xi_0$, the transformation corresponds to time reversal of the classical solutions (see \cref{equ:relcl}). It may thus not be surprising that when looking at the expectation values of $\xi$ we will see that the sign of $\beta$ will correspond to the sign of the mass of the black/white hole, analogous to $\xi_0$ is the classical theory. Requiring the wavefunction to respect this symmetry ensures that time-reversed classical solutions, which relate black and white hole solutions, are treated equivalently.

It is instructive to write the general state subject to \cref{equ:alphacon} slightly differently than before:
\begin{align}
    \psi = \int_0^{\infty}\frac{\td k}{2\pi}\int_{-\infty}^{\infty}\frac{\td\Lambda}{2\pi}\,\alpha(\Lambda,k)e^{\im\Lambda T}\left[e^{\im\left[k\xi+\frac{1}{k}\left(\frac{\Lambda}{3}\eta^3-\eta\right)\right]}+e^{\im\chi(k)}e^{-\im\left[k\xi+\frac{1}{k}\left(\frac{\Lambda}{3}\eta^3-\eta\right)\right]}\right]\,.
\end{align}
Requiring that the solution for given $\Lambda$ and $k$, i.e., the sum of the two terms inside the square brackets, is invariant under (\ref{equ:symham}) fixes the free function $\chi(k)$ to $\chi(k) = k\beta$.
After restricting ourselves to this specific choice, an alternative way of writing the allowed states is as
\begin{equation}
\psi(\eta,\xi,T) = \int_0^{\infty}\frac{\td k}{2\pi}\int_{-\infty}^{\infty}\frac{\td\Lambda}{2\pi}\,\tilde{\alpha}(\Lambda,k)e^{\im\Lambda T}\cos\left(k\left(\xi-\frac{\beta}{2}\right)+\frac{1}{k}\left(\frac{\Lambda}{3}\eta^3-\eta\right)\right)\,, 
\label{eq:cossolution}
\end{equation} 
where a phase has been absorbed into $\tilde\alpha(\Lambda,k)$. This expression explicitly shows that the general allowed solution for given $\Lambda$ is built from generalised real combinations of the  initially independent solutions to the Wheeler--DeWitt equation. We can then also justify why we called the condition on $\alpha$ given in (\ref{equ:alphacon}) a boundary condition: on the boundary of configuration space given by $\eta=0$ we can see that (\ref{eq:cossolution}) is a linear combination of functions $\cos(k(\xi-\frac{\beta}{2}))$ which are even under reflection about $\frac{\beta}{2}$. We hence have $\psi(0,\xi,T) = \psi(0,\beta-\xi,T)$, and these particular self-adjoint extensions are analogous to the ones we constructed in section \ref{sec:interior} for the theory restricted to black/white hole interiors, given that they also arise from symmetries of the wavefunction on the boundary of configuration space.

Now that we have systematically derived all possible self-adjoint extensions for $\xi\in\mathbb{R}$ and restricted them to a one-parameter family of physically motivated quantum theories, we will analyse these theories in the next section by calculating semiclassical expectation values. Importantly we will find that a quantum theory either only includes semiclassical states that correspond to classical solutions with positive mass, only semiclassical states corresponding to negative mass, or only semiclassical states corresponding to zero mass. In other words, two semiclassical states of the same quantum theory always have the same sign for their black/white hole mass. The zero mass quantum theory will be considered separately. We will also look at states without a semiclassical interpretation and find tunnelling states. 

\section{Analysing the quantum theory}
\label{sec:aqt}

In this section we define a semiclassical state and use it to compute the expectation values of the operators $\hat{\eta}$ and $\hat{\xi}$, which are the quantum analogues of components of the spacetime metric as defined in \cref{equ:lineel}. Since expectation values alone do not fully capture the characteristics of a quantum state, we also calculate the variance of $\hat{\xi}$ to verify that our state remains semiclassical. Additionally, we will examine the self-adjoint extension with $\beta = 0$, which after identification of $\beta$ with the classical parameter $\xi_0$ corresponds to a spacetime without a black or white hole. Finally, we will consider a class of highly quantum states that describe a tunnelling spacetime.

Any state in our theory is characterised by a choice of $\alpha(\Lambda,k)$ in \cref{equ:generalwf}. A seemingly natural choice for a semiclassical state would be a Gaussian state 
\begin{align}
    \alpha_{sc}(\Lambda,k) = Ne^{-\frac{(k-k_c)^2}{2\sigma_k^2}}e^{-\frac{(\Lambda-\Lambda_c)^2}{2\sigma_{\Lambda}^2}}
\end{align}
peaked on some classical values $k_c$ and $\Lambda_c$, where $N$ is a normalisation factor. However it is clear that this does not satisfy the boundary condition in \cref{equ:alphacon}. Thus we will consider the following state:
\begin{align}
\label{equ:alphasc}
    \alpha_{sc}(\Lambda,k) = N e^{-\im\frac{\beta}{2} k} e^{-\frac{(|k|-k_c)^2}{2\sigma_k^2}}e^{-\frac{(\Lambda-\Lambda_c)^2}{2\sigma_{\Lambda}^2}}\,.
\end{align}
For a given choice of $\beta$, this $\alpha$ function defines a state in the quantum theory with self-adjoint extension parameter $\beta$. Notice that the Gaussian now has peaks at $k=\pm k_c$, where we choose $k_c>0$. We will show that this state is semiclassical in the next section by calculating its variance, due to the absolute value of $k$ in the Gaussian this is not obvious. Using the norm (\ref{eq:normexpr}), the normalisation factor $N$ for this state can be computed as 
\begin{align}
    1 \stackrel{!}{=} \langle\psi_{sc},\psi_{sc}\rangle &= N^2\int_{-\infty}^{\infty}\frac{\td k}{2\pi}\int_{-\infty}^{\infty}\frac{\td \Lambda}{2\pi}|k| e^{-\frac{(|k|-k_c)^2}{\sigma_k^2}}e^{-\frac{(\Lambda-\Lambda_c)^2}{\sigma_{\Lambda}^2}}\\
    &= N^2 \frac{\sigma_{\Lambda}}{2\sqrt{\pi}}\int_{-k_c}^{\infty}\frac{\td q}{2\pi}\,2(q+k_c)e^{-\frac{q^2}{\sigma_k^2}}\\
    &= N^2 \frac{\sigma_{\Lambda}}{2\sqrt{\pi}}\left[\frac{\sigma_k^2}{2\pi}e^{-\frac{k_c^2}{\sigma_k^2}}+\frac{k_c\sigma_k}{2\sqrt{\pi}}\,\text{erfc}\left(-\frac{k_c}{\sigma_k}\right)\right]
\end{align}
so that
\begin{equation}
    N^2 = \frac{2\sqrt{\pi}}{\sigma_{\Lambda}}\left[\frac{\sigma_k^2}{2\pi}e^{-\frac{k_c^2}{\sigma_k^2}}+\frac{k_c\sigma_k}{2\sqrt{\pi}}\,\text{erfc}\left(-\frac{k_c}{\sigma_k}\right)\right]^{-1}\,.
\end{equation}
Here erfc denotes the complementary error function. The semiclassical state is defined by specifying $\Lambda_c$, $k_c$, $\sigma_{\Lambda}$, $\sigma_k$ and $\beta$. We will now use this class of states to analyse the semiclassical regime of quantum theories defined for different $\beta$.

\subsection{Semiclassical expectation values}
First we will look at the expectation value of $\hat{\eta}$. The calculation given in \cref{sec:Aev} shows that
\begin{align}
     \begin{split}
     \langle\hat{\eta}\rangle_{sc} = \langle \psi_{sc},\hat\eta\psi_{sc}\rangle = & 
    \,N^2\frac{\sqrt[3]{3\sL^2\sk^7}\Gamma\left(\frac{2}{3}\right)}{4\pi^2}{}_1F_1\left(-\frac{1}{6},\frac{1}{2},-\sigma_{\Lambda}^2T^2\right)\\
    &\times\left[\Gamma\left(\frac{7}{6}\right){}_1F_1\left(-\frac{2}{3},\frac{1}{2},-\frac{k_c^2}{\sigma_k^2}\right)+2\Gamma\left(\frac{5}{3}\right)\frac{k_c}{\sigma_k}{}_1F_1\left(-\frac{1}{6},\frac{3}{2},-\frac{k_c^2}{\sigma_k^2}\right)\right]
    \end{split}
    \label{eq:etaexpv}
\end{align}
where ${}_1F_1$ denotes the confluent hypergeometric function. One can immediately notice that $\langle\hat{\eta}\rangle_{sc}$ is independent of $\Lambda_c$, just as the classical solution $\eta(T)=\sqrt[3]{3kT}$ (given in \cref{equ:relcl}) is independent of $\Lambda$. $\langle\hat{\eta}\rangle_{sc}$ does however depend on the variance $\sigma_\Lambda$, which is expected as the unimodular time $T$ is conjugate to $\Lambda$ and therefore larger variance in the cosmological constant is expected to result in closer agreement with the classical solutions. The expectation value is also independent of the self-adjoint extension parameter $\beta$. Again this can already be seen in the classical solution, which does not depend on $\xi_0$. Before plotting this solution, one can already see that the asymptotic behaviour of the expectation value at late or early times agrees with the classical solution. The asymptotic expansion for the confluent hypergeometric function is given by \citep{abramowitz_handbook_1964}
\begin{align}
    {}_1F_1(a,b,z) \sim \Gamma(b)\left(\frac{e^zz^{a-b}}{\Gamma(a)}+\frac{(-z)^{-a}}{\Gamma(b-a)}\right)\,.
\end{align}

If we also assume a semiclassical state for which $\sigma_k \ll k_c$, we can approximate $N^2\sim 2\pi/(k_c \sigma_k \sigma_\Lambda)$ and using the asymptotic forms of the hypergeometric function yields 
\begin{align}
    \langle\hat{\eta}\rangle_{sc} &\sim \sqrt[3]{3k_c|T|}\,,\quad |T|\rightarrow\infty\,.
    \label{eq:etaasympt}
\end{align}
We thus see that for any $\sL$ the expectation value will asymptotically behave like a classical solution with $k=+k_c$ as $T\rightarrow +\infty$, and like a classical solution with $k=-k_c$ as $T\rightarrow -\infty$. This result already indicates that our semiclassical state describes a transition from an asymptotic black hole to an asymptotic white hole solution.

\begin{figure}
     \centering
     \begin{subfigure}[b]{0.48\textwidth}
         \centering
         \includegraphics[width=\textwidth]{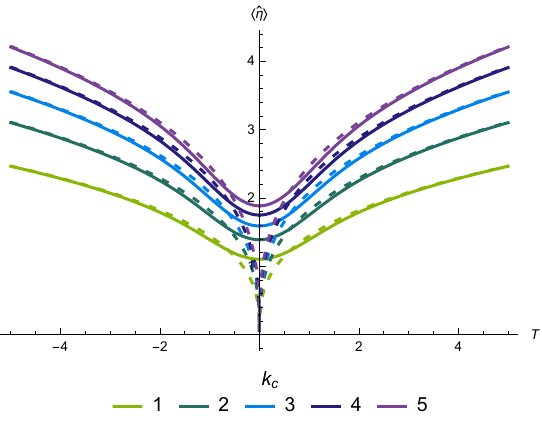}
         \caption{$\langle\hat{\eta}\rangle_{sc}$ for $\sL = 1$, $\sk = 0.1$ and various $k_c$. (The solution is indepedent of $\Lambda_c$.)}
     \end{subfigure}
     \hfill
     \begin{subfigure}[b]{0.48\textwidth}
         \centering
         \includegraphics[width=\textwidth]{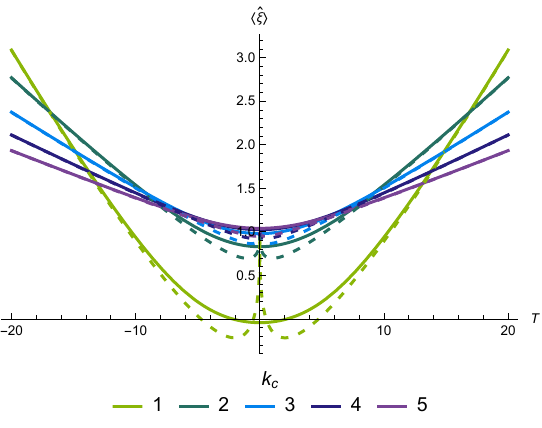}
         \caption{$\langle\hat{\xi}\rangle_{sc}$ for $\Lambda_c = 0.3$, $\sL = 0.2$, $\sk = 0.1$, $\beta = 2$ and various $k_c$.}
     \end{subfigure}
     \hfill
     \begin{subfigure}[b]{0.48\textwidth}
         \centering
         \includegraphics[width=\textwidth]{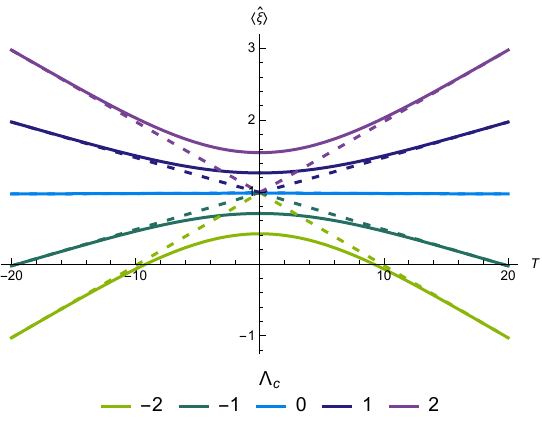}
         \caption{$\langle\hat{\xi}\rangle_{sc}$ for $k_c = 20$, $\sk = 0.1$, $\sL = 0.1$, $\beta = 2$ and various $\Lambda_c$.}
         \label{fig:qgxiL}
     \end{subfigure}
        \caption{Visualisation of the expectation values $\langle\hat{\eta}\rangle_{sc}$ and $\langle\hat{\xi}\rangle_{sc}$ for different values of $k_c$ and $\Lambda_c$. The solid lines correspond to the quantum expectation values and the dashed lines of the same colour correspond to the classical solutions with $k=\pm k_c$ and $\Lambda=\Lambda_c$. Here $\beta = 2$. One can see that for large $T$ the classical and quantum solutions agree very well but that $\langle\hat{\eta}\rangle_{sc}$ does not vanish at the origin, resolving the singularity.}
        \label{fig:qgsoletaxi}
\end{figure}

The expectation value of $\hat{\xi}$ can be calculated in a similar way, as is shown in \cref{sec:Aev}. The result is:
\begin{align}
    \begin{split}
    \langle\hat\xi\rangle_{sc} 
    = \langle \psi_{sc},\hat\xi\psi_{sc}\rangle =& \,\frac{N^2}{2\pi}\left[\frac{\Lambda_c}{2}\sigma_k\text{erfc}\left(-\frac{k_c}{\sigma_k}\right)\left(\frac{1}{\sqrt{\pi}}e^{-\sigma_{\Lambda}^2T^2}+\sL T\text{erf}(\sL T)\right)\right.\\
    &\left.-\frac{1}{\pi}\Gamma\left(\frac{2}{3}\right)\sqrt[3]{3\sL^2\sk}\left(3\Gamma\left(\frac{7}{6}\right){}_1F_1\left(\frac{1}{3},\frac{1}{2},-\frac{k_c^2}{\sk^2}\right)\right.\right.\\
    &\left.\left.+\Gamma\left(\frac{2}{3}\right)\frac{k_c}{\sk}{}_1F_1\left(\frac{5}{6},\frac{3}{2},-\frac{k_c^2}{\sk^2}\right)\right){}_1F_1\left(-\frac{1}{6},\frac{1}{2},-\sL^2 T^2\right)\right]+\frac{\beta}{2}\,.
    \end{split}
    \label{eq:xiexpv}
\end{align}
Here erfc denotes the complementary error function and erf the error function. If we look at the asymptotic behaviour of this expression for large $|T|$ and $\sigma_k \ll k_c$, we again find agreement with classical solutions:
\begin{align}
    \langle\xi(T)\rangle_{sc}\sim \frac{\Lambda_c}{k_c}|T|-\sqrt[3]{\frac{3|T|}{k_c^5}}+\frac{\beta}{2}\,,\quad |T|\rightarrow\infty\,.
\end{align}
Again, we see that this solution interpolates between an asymptotic classical solution (\ref{equ:relcl}) for $k=-k_c$ at large negative times and an asymptotic classical solution for $k=+k_c$ at large positive times. Moreover, the self-adjoint extension parameter $\beta$ acts as a constant shift in $\langle\hat\xi\rangle_{sc}$ analogous to the constant shift $\xi_0$ of the classical solution, with the precise relation $\beta=2\xi_0$ expected from  the discussion below \cref{equ:symham}. Furthermore, if we recall the classical definition of the black/white hole mass as $M = 4\pi k^2\xi_0$, we see that the sign of $\beta$ determines the sign of the mass of the black/white hole in the classical limit. But $\beta$ is the self-adjoint extension parameter, thus changing $\beta$ means changing the quantum theory: a given quantum theory (with fixed $\beta$) only has semiclassical black hole states with $\text{sgn}(M) = \text{sgn}(\beta)$. By choosing a quantum theory with positive $\beta$ we can exclude negative mass black/white hole states from our theory and thus do not run into the problem of a vacuum instability. Plots of $\langle\hat{\eta}\rangle_{sc}$ and $\langle\hat{\xi}\rangle_{sc}$ are shown in \cref{fig:qgsoletaxi}. The functions are defined for negative as well as positive time $T$, as the expectation value of $\hat{\eta}$ is positive everywhere. Since $\langle\hat{\eta}\rangle_{sc}$ remains finite at the origin one sees that the singularity is resolved. The resulting spacetime starts as a classical black hole spacetime at large negative times, undergoes a bounce, and evolves into a classical white hole spacetime at late times, retaining the same mass as the initial black hole.

\begin{figure}
     \centering
     \begin{subfigure}[b]{0.48\textwidth}
         \centering
         \includegraphics[width=\textwidth]{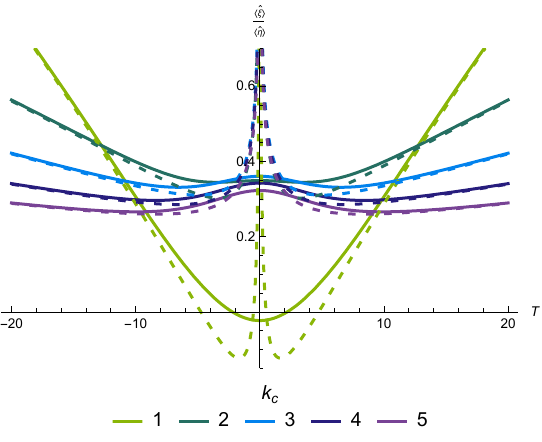}
         \caption{$\langle\hat{\xi}\rangle_{sc}/\langle\hat{\eta}\rangle_{sc}$ for $\Lambda_c = 0.3$, $\sL = 0.2$, $\sk = 0.1$, $\beta = 2$ and various $k_c$.}
     \end{subfigure}
     \hfill
     \begin{subfigure}[b]{0.48\textwidth}
         \centering
         \includegraphics[width=\textwidth]{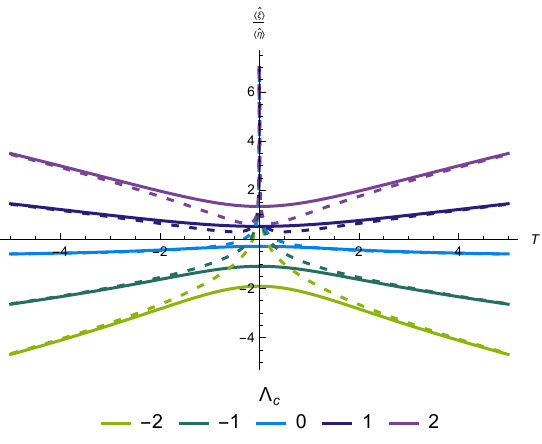}
         \caption{$\langle\hat{\xi}\rangle_{sc}/\langle\hat{\eta}\rangle_{sc}$ for $k_c = 1$, $\sL = 0.5$, $\sk = 0.1$, $\beta = 2$ and various $\Lambda_c$.}
     \end{subfigure}
        \caption{The solid lines show $\langle\hat{\xi}\rangle_{sc}/\langle\hat{\eta}\rangle_{sc}$, while the dashed lines in the same colour correspond to  classical solutions with $\Lambda=\Lambda_c$ and $k=\pm k_c$. For a semiclassical state, these plots of $\langle\hat{\xi}\rangle_{sc}/\langle\hat{\eta}\rangle_{sc}$ approximately describe the behaviour of $\langle\widehat{\xi/\eta}\rangle_{sc}$ and therefore show that the singularity at the origin is resolved in the quantum theory.}
        \label{fig:qgsolsfsc}
\end{figure}

\begin{figure}
     \centering
     \begin{subfigure}[b]{0.48\textwidth}
         \centering
         \includegraphics[width=\textwidth]{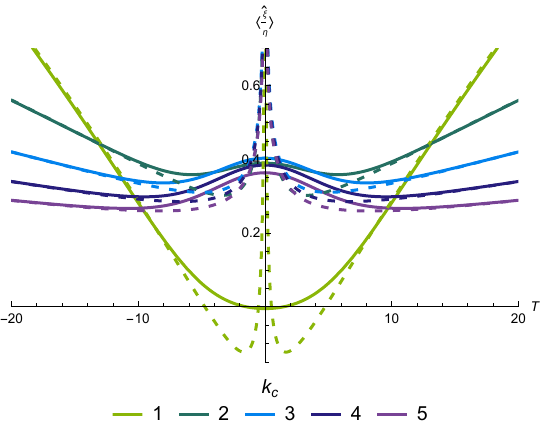}
         \caption{$\langle\widehat{\xi/\eta}\rangle_{sc}$ for $\Lambda_c = 0.3$, $\sL = 0.2$, $\sk = 0.1$, $\beta = 2$ and various $k_c$.}
         \label{fig:sfev1}
     \end{subfigure}
     \hfill
     \begin{subfigure}[b]{0.48\textwidth}
         \centering
         \includegraphics[width=\textwidth]{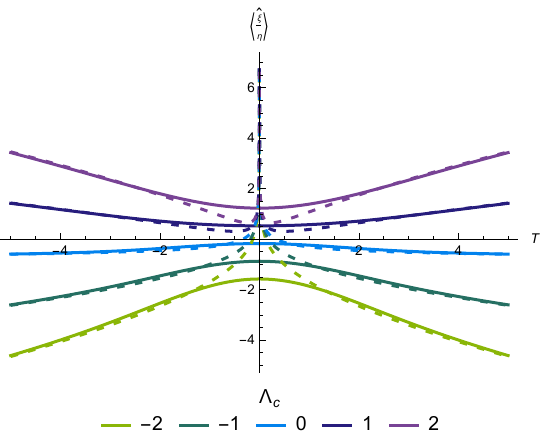}
         \caption{$\langle\widehat{\xi/\eta}\rangle_{sc}$ for $k_c = 1$, $\sL = 0.5$, $\sk = 0.1$, $\beta = 2$ and various $\Lambda_c$.}
         \label{fig:sfev2}
     \end{subfigure}
     \hfill
     \begin{subfigure}[b]{0.48\textwidth}
         \centering
         \includegraphics[width=\textwidth]{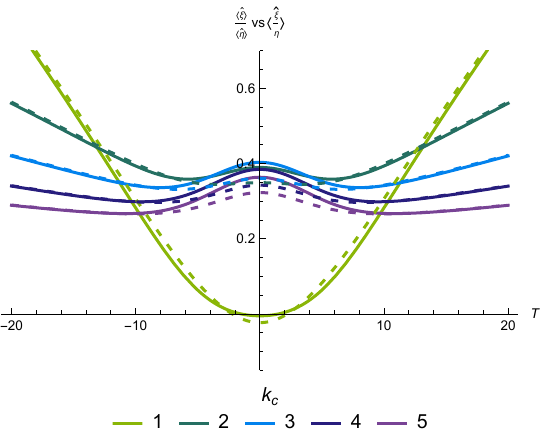}
         \caption{$\langle\widehat{\xi/\eta}\rangle_{sc}$ vs. $\langle\hat{\xi}\rangle_{sc}/\langle\hat{\eta}\rangle_{sc}$ for $\Lambda_c = 0.3$, $\sL = 0.2$, $\sk = 0.1$, $\beta = 2$ and various $k_c$.}
        \label{fig:sfvsxidiveta}
     \end{subfigure}
        \caption{The solid lines show $\langle\widehat{\xi/\eta}\rangle_{sc}$, while the dashed lines in \cref{fig:sfev1} and \cref{fig:sfev2} in the same colour correspond to the classical solutions with $\Lambda=\Lambda_c$ and $k=\pm k_c$. In \cref{fig:sfvsxidiveta} the dashed lines correspond to $\langle\hat{\xi}\rangle_{sc}/\langle\hat{\eta}\rangle_{sc}$. The plots look very similar to each other indicating that the state is indeed semiclassical. We can again see that there is no divergence of metric components in the quantum theory, meaning that the singularity is resolved.}
        \label{fig:qgsolsf}
\end{figure}

Another interesting expectation value to look at is that of the operator $\widehat{\xi/\eta}$, as this corresponds to the metric component  $g_{zz}$ given in \cref{equ:lineel}, which diverges classically at the singularity. Before calculating this expectation value we can look at the ratio $\langle\hat{\xi}\rangle_{sc}/\langle\hat{\eta}\rangle_{sc}$ of the expectation values we have already calculated; if our assumption of semiclassicality of the state is correct this should give similar results to the full quantum expectation value $\langle\widehat{\xi/\eta}\rangle_{sc}$ which we might see as the direct analogue of $g_{zz}$ in our quantum theory. As one can see in \cref{fig:qgsolsfsc} there are no singularities, which follows from the fact that $\langle\hat{\eta}\rangle_{sc}$ remains bounded away from zero during the quantum evolution.
Calculating $\langle\widehat{\xi/\eta}\rangle_{sc}$ shows that it agrees very well with the approximation above. Details of the calculation can be seen in \cref{sec:Aev}, here we simply state the result:
\begin{align}
    \begin{split}
    \langle\widehat{\xi/\eta}\rangle_{sc} 
    =& \, \frac{N^2}{4\pi^2}\left[\beta\sqrt[3]{\frac{\sL^4\sk^5}{3}}{}_1F_1\left(\frac{1}{6},\frac{1}{2},-\sL^2T^2\right)\left(\frac{\sqrt{\pi}}{2^{2/3}}\Gamma\left(\frac{2}{3}\right){}_1F_1\left(-\frac{1}{3},\frac{1}{2},-\frac{k_c^2}{\sk^2}\right)\right.\right.\\
    &\left.\left.+3\frac{k_c}{\sk}\Gamma\left(\frac{4}{3}\right)^2{}_1F_1\left(\frac{1}{6},\frac{3}{2},-\frac{k_c^2}{\sk^2}\right)\right)+\Lambda_c\Gamma\left(\frac{5}{6}\right)\sqrt[3]{\frac{\sL\sk^2}{3}}{}_1F_1\left(-\frac{1}{3},\frac{1}{2},-\sL^2T^2\right)\right.\\
    &\left.\times\left(\Gamma\left(\frac{1}{3}\right){}_1F_1\left(\frac{1}{6},\frac{1}{2},-\frac{k_c^2}{\sk^2}\right)+2\frac{k_c}{\sk}\Gamma\left(\frac{5}{6}\right){}_1F_1\left(\frac{2}{3},\frac{3}{2},-\frac{k_c^2}{\sk^2}\right)\right)\right.\\
    &\left.-\sqrt{\pi}\sL\intinf\td k\frac{1}{|k|}\left(e^{-\frac{(|k|-k_c)^2}{\sk^2}}-e^{-\frac{k_c^2}{\sk^2}}\right)\right]\,.
    \end{split}
\end{align}
The integral in the last line has to be integrated numerically. The function is plotted in \cref{fig:qgsolsf} for the same parameter values previously chosen in \cref{fig:qgsolsfsc}. We see that the general behaviour of the full expectation value $\langle\widehat{\xi/\eta}\rangle_{sc}$ is very similar to that of the ratio $\langle\hat{\xi}\rangle_{sc}/\langle\hat{\eta}\rangle_{sc}$, and in particular a resolution of the classical singularity is also seen in $\langle\widehat{\xi/\eta}\rangle_{sc}$. An explicit comparison of the two in \cref{fig:sfvsxidiveta}  shows quantitative differences especially near $T=0$ but very similar qualitative behaviour.

\newpage To conclude this subsection let us stress again that (\ref{eq:etaexpv}) and (\ref{eq:xiexpv}) are analytical expressions for what we might consider to be the quantum-corrected Schwarzschild--(Anti-)de Sitter metric in our theory. That is, we could write down a semiclassical line element (now expressed in unimodular time $T$)
\begin{align}
    \td s^2 = -\frac{\td T^2}{\langle\hat\eta(T)\rangle^3_{sc}\langle\hat\xi(T)\rangle_{sc}} + \frac{\langle\hat\xi(T)\rangle_{sc}}{\langle\hat\eta(T)\rangle_{sc}} \td z^2 + \langle\hat{\eta}(T)\rangle_{sc}^2\td\Omega^2
\end{align}
which describes the nonsingular geometry obtained from incorporating quantum corrections at high curvature. This result is exact within the minisuperspace framework we are using, but the resulting expressions are rather unwieldy. Recall that our expectation values depend on the four free parameters in the state, namely $\Lambda_c$, $k_c$, $\sigma_\Lambda$, $\sigma_k$, and the self-adjoint extension parameter $\beta$. A useful approximation would be to assume that $\sigma_k\ll k_c$, i.e., the state is sharply peaked in $k$, as we already did before \cref{eq:etaasympt}. This allows replacing some of the hypergeometric functions with their asymptotic expressions and we have
\begin{align}
\langle\hat\eta(T)\rangle_{sc} & \approx \frac{1}{\sqrt{\pi}}\left(\frac{3 k_c}{\sigma_\Lambda}\right)^{1/3}\Gamma\left(\frac{2}{3}\right){}_1F_1\left(-\frac{1}{6},\frac{1}{2},-\sL^2T^2\right)
\,,
\\ \langle\hat\xi(T)\rangle_{sc} & \approx \frac{\Lambda_c}{\sqrt{\pi}\sigma_\Lambda k_c}e^{-\sigma_\Lambda^2 T^2} + \frac{\Lambda_c}{k_c} T\,{\rm erf}(\sigma_\Lambda T) +\frac{\Gamma\left(-\frac{1}{3}\right){}_1F_1\left(-\frac{1}{6},\frac{1}{2},-\sL^2T^2\right)}{\sqrt{\pi}\sqrt[3]{9 k_c^5 \sigma_\Lambda}} + \frac{\beta}{2}
\,.
\end{align}
These provide excellent approximations to the full quantum solution and still show nonsingular behaviour with an asymptotic transition between black and white hole.

\subsection{Zero mass self-adjoint extension}
\begin{figure}
     \centering
     \begin{subfigure}[b]{0.48\textwidth}
         \centering
         \includegraphics[width=\textwidth]{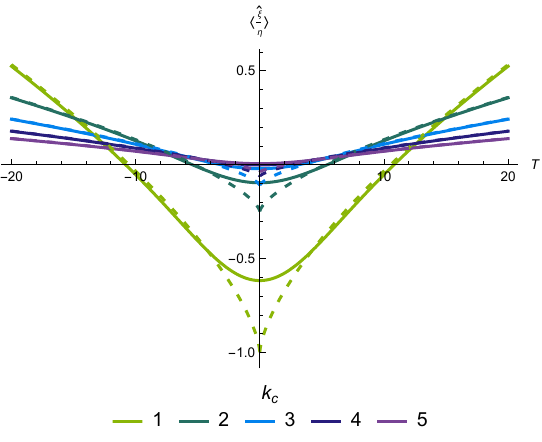}
         \caption{$\langle\widehat{\xi/\eta}\rangle_{sc}$ for $\Lambda_c = 0.3$, $\sL = 0.2$, $\sk = 0.1$, $\beta = 0$ and various $k_c$.}
     \end{subfigure}
     \hfill
     \begin{subfigure}[b]{0.48\textwidth}
         \centering
         \includegraphics[width=\textwidth]{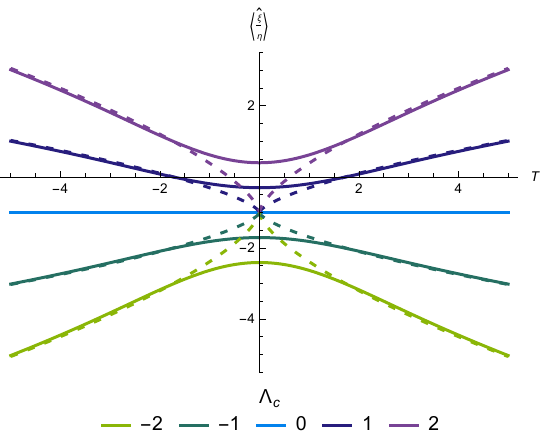}
         \caption{$\langle\widehat{\xi/\eta}\rangle_{sc}$ for $k_c = 1$, $\sL = 0.5$, $\sk = 0.1$, $\beta = 0$ and various $\Lambda_c$.}
     \end{subfigure}
        \caption{The solid lines show $\langle\widehat{\xi/\eta}\rangle_{sc}$, while the dashed lines in the same colour correspond to the classical solutions with $\Lambda=\Lambda_c$ and $k=k_c$. One can see that the classical solutions do not diverge for $T = 0$, as there is no black or white hole.}
        \label{fig:sfAdS}
\end{figure}

An interesting special case is that of the self-adjoint extension $\beta = 0$, which would correspond to the classical parameter $\xi_0=0$, i.e., an (Anti-)de Sitter spacetime without a black or white hole. This case is interesting because here the classical theory does not have a curvature singularity for $T = 0$; instead there is a coordinate singularity similar to the one in polar coordinates at the origin. The quantum theory cannot distinguish between coordinate and curvature singularities (as also discussed in \cite{gielen_unitarity_2022-1,gielen_black_2024,gielen_neves} in similar contexts) and thus we still find ``singularity resolution'' in the sense of strong departures from the classical solution in this case, as can be seen in \cref{fig:sfAdS}. 
The meaning of the part of the solution parametrised by negative $T$ can be illustrated in a conformal diagram as shown in \cref{fig:dScondiag} for de Sitter spacetime ($\Lambda>0$). The left square describes a de Sitter spacetime of our classical theory. The dark grey region is the static patch of de Sitter that is covered by the coordinates we used. Since $\xi$ is not constrained to negative values, the cosmological horizon at the edge of the static patch is not a boundary of our minisuperspace. In the case of negative $k$, $\xi>0$ covers the past triangle, region IVa in the conformal diagram. If instead $k>0$, the part of our solution with $\xi>0$ covers the future triangle, region IIa. The edge denoted with $T = 0$ is the end of de Sitter space. What we see in the quantum theory is that the solutions are extended beyond de Sitter into a new de Sitter spacetime, the square on the right. Asymptotically for large $T$ the solution again describes the static patch of a de Sitter universe, here the light grey shaded region. The chequered region between the two de Sitter spacetimes illustrates the quantum region, where a classical description of spacetime is not possible. The quantum solutions shown in \cref{fig:sfAdS} (those with $\Lambda_c >0$) cover regions IVa, Ia, Ib and IIb.
\begin{figure}
\begin{center}
    \begin{tikzpicture}
        \colorlet{mydarkpurple}{blue!40!red!50!black}
        \colorlet{mylightpurple}{mydarkpurple!80!red!6}

        \node (I) at (-3.5,0) {};
        \node (II) at (3.5,0) {};
        \node (Itl) at (-6.5,3) {};
        \node (Ibl) at (-6.5,-3) {};
        \node (Ibr) at (-0.5,-3) {};
        \node (Itr) at (-0.5,3) {};
        \node (IItl) at (0.5,3) {};
        \node (IIbl) at (0.5,-3) {};
        \node (IItr) at (6.5,3) {};
        \node (IIbr) at (6.5,-3) {};
        \draw (Ibl.center) -- (IIbr.center) -- (IItr.center) -- (Itl.center) -- (Ibl.center);
        \draw (Ibl.center) -- (I.center) -- (Itl.center);
        \draw[fill=gray!20] (IItl.center) -- (II.center) -- (IIbl.center) -- (IItl.center) -- cycle;
        \draw[pattern=checkerboard, pattern color = gray!20] (Itr.center) -- (IItl.center) -- (IIbl.center) -- (Ibr.center) -- (Itr.center) -- cycle;
        \draw[fill=gray!70]   (Ibr.center) --
                    node[midway,below,sloped] {horizon}
                (I.center) --
                    node[midway,above,sloped] {horizon}
                (Itr.center) --
                    node[midway,above,sloped] {$T = 0$}
                (Ibr.center) -- cycle;
        \draw   (IItr.center) -- (IIbl.center);
        \draw   (IItl.center) -- (IIbr.center);

        \tikzset{declare function={%
            kruskal(\x,\c)  = {\fpeval{asin( \c*sin(2*\x) )*2/pi}};
        }}

        \draw[blue!40!red!80!black,line width=0.4,samples=20,smooth,variable=\x,domain=-6.5:-0.5] 
                plot(\x,{-4*kruskal((\x+3.5)*pi/16,0.5)})
                plot(\x,{ 4*kruskal((\x+3.5)*pi/16,0.5)});

        \draw[blue!40!red!80!black,line width=0.4,samples=20,smooth,variable=\y,domain=-3:3] 
                plot({-4*kruskal(\y*pi/16,0.5)-3.5},\y)
                plot({ 4*kruskal(\y*pi/16,0.5)-3.5},\y);

        \draw[blue!40!red!80!black,line width=0.4,samples=20,smooth,variable=\x,domain=0.5:6.5] 
                plot(\x,{-4*kruskal((\x-3.5)*pi/16,0.5)})
                plot(\x,{ 4*kruskal((\x-3.5)*pi/16,0.5)});

        \draw[blue!40!red!80!black,line width=0.4,samples=20,smooth,variable=\y,domain=-3:3] 
                plot({-4*kruskal(\y*pi/16,0.5)+3.5},\y)
                plot({ 4*kruskal(\y*pi/16,0.5)+3.5},\y);

        \coordinate (IIIat) at (-6.5,1.22277);
        \coordinate (IIIab) at (-6.5,-1.22277);
        \coordinate (Iat) at (-0.5,1.22277);
        \coordinate (Iab) at (-0.5,-1.22277);
        \coordinate (IIal) at (-4.72277,3);
        \coordinate (IIar) at (-2.27723,3);
        \coordinate (IVal) at (-4.72277,-3);
        \coordinate (IVar) at (-2.27723,-3);
        \coordinate (IIIbt) at (6.5,1.22277);
        \coordinate (IIIbb) at (6.5,-1.22277);
        \coordinate (Ibt) at (0.5,1.22277);
        \coordinate (Ibb) at (0.5,-1.22277);
        \coordinate (IIbr) at (4.72277,3);
        \coordinate (IIbl) at (2.27723,3);
        \coordinate (IVbr) at (4.72277,-3);
        \coordinate (IVbl) at (2.27723,-3);

        \fill[fill=mylightpurple, opacity=0.7]
            plot[smooth,domain=-6.5:-0.5,variable=\x] 
            (\x,{ 4*kruskal((\x+3.5)*pi/16,0.5)})
            -- (Iat) -- (Iab) --
            plot[smooth,domain=0.5:6.5,variable=\x] 
            (-\x,{-4*kruskal((-\x+3.5)*pi/16,0.5)})
            -- (IIIat) -- (IIIab)
            -- cycle;

        \fill[fill=mylightpurple, opacity=0.7]
            plot[smooth,domain=-3:3,variable=\y] 
            ({ 4*kruskal(\y*pi/16,0.5)-3.5},\y)
            -- (IIar) -- (IIal) --
            plot[smooth,domain=-3:3,variable=\y] 
            ({-4*kruskal(\y*pi/16,0.5)-3.5},\y)
            -- (IVar) -- (IVal)
            -- cycle;

        \fill[fill=mylightpurple, opacity=0.7]
            plot[smooth,domain=0.5:6.5,variable=\x] 
            (\x,{ 4*kruskal((\x-3.5)*pi/16,0.5)})
            -- (IIIbt) -- (IIIbb) --
            plot[smooth,domain=0.5:6.5,variable=\x] 
            (\x,{-4*kruskal((\x-3.5)*pi/16,0.5)})
            -- (Ibt) -- (Ibb)
            -- cycle;

        \fill[fill=mylightpurple, opacity=0.7]
            plot[smooth,domain=-3:3,variable=\y] 
            ({ 4*kruskal(\y*pi/16,0.5)+3.5},\y)
            -- (IIbr) -- (IIbl) --
            plot[smooth,domain=-3:3,variable=\y] 
            ({-4*kruskal(\y*pi/16,0.5)+3.5},\y)
            -- (IVbr) -- (IVbl)
            -- cycle;
        
        \node (Ia) at (-1.5,0) {Ia};
        \node (IIa) at (-3.5,1.5) {IIa};
        \node (IIIa) at (-5,0) {IIIa};
        \node (IVa) at (-3.5,-1.5) {IVa};
        \node (Ib) at (1.5,0) {Ib};
        \node (IIb) at (3.5,1.5) {IIb};
        \node (IIIb) at (5,0) {IIIb};
        \node (IVb) at (3.5,-1.5) {IVb};
    \end{tikzpicture}
    \end{center}
\caption{Conformal diagram of two de Sitter spacetimes connected by a "quantum region". The static patch of both spacetimes is shaded in grey. The purple shaded regions indicate the patches for which $z\in I$, where $I$ is our compactification interval.}
    \label{fig:dScondiag}
\end{figure}

If one chooses a self-adjoint extension with $\beta\neq 0$ one ensures that the semiclassical states correspond to black/white hole spacetimes and the singularities that are resolved are curvature singularities, not just coordinate singularities. In this sense the $\beta>0$ theories seem to lead to physically more sensible notions of singularity resolution. The case of a $\beta<0$ self-adjoint extension would again resolve curvature singularities, but these are the timelike naked singularities of classical relativity that we would perhaps rather not resolve, as we discussed in the introduction.

\subsection{Semiclassicality and tunnelling states}
Since we have reduced the system to two degrees of freedom using symmetry arguments before quantisation, our results are primarily applicable at the semiclassical level; a minisuperspace approach such as ours excludes a quantum-level relevance of inhomogeneous degrees of freedom, an assumption we cannot rigorously justify. We anticipate that a full quantum theory of gravity (defined outside of a formal canonical quantisation, which cannot be made rigorous) could significantly alter our results in the highly quantum regime. 

In a previous section we have seen that $\langle \hat{\xi}\rangle_{sc}/\langle \hat{\eta}\rangle_{sc}$ and $\langle \hat{\xi}/\hat{\eta}\rangle_{sc}$ agree very well, which is a good indication that the absolute value of the $k$ parameter in the Gaussian of our state defined in \cref{equ:alphasc} does not ruin the good semiclassicality properties of the state. Here we will explicitly calculate the variance of $\hat{\xi}$, $\text{Var}(\xi)=\sqrt{\langle\hat\xi^2\rangle-\langle\hat\xi\rangle^2}$, in the state $\psi_{sc}$ and see by comparing it to the expectation value that the state is indeed semiclassical everywhere.

\begin{figure}
     \centering
     \begin{subfigure}[b]{0.48\textwidth}
         \centering
         \includegraphics[width=\textwidth]{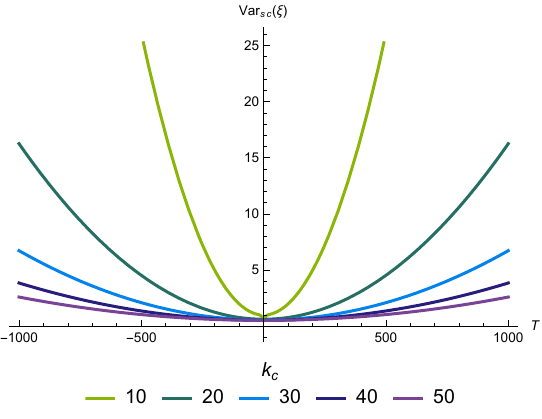}
         \caption{$\text{Var}_{sc}(\xi)$ for $\Lambda_c = 1$, $\sL = 0.1$, $\sk = 1$.}
         \label{fig:varxisck}
     \end{subfigure}
     \hfill
     \begin{subfigure}[b]{0.48\textwidth}
         \centering
         \includegraphics[width=\textwidth]{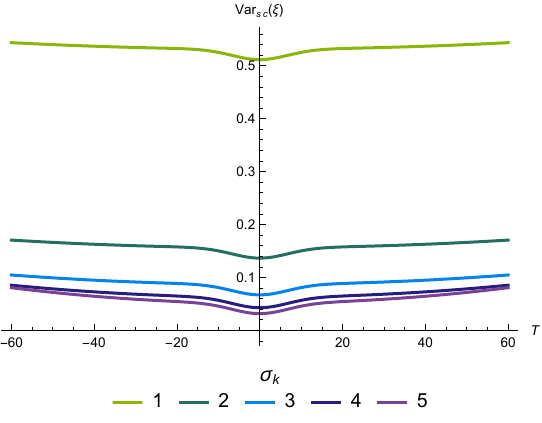}
         \caption{$\text{Var}_{sc}(\xi)$ for $\Lambda_c = 1$, $\sL = 0.1$, $k_c = 40$.}
         \label{fig:varxiscsk}
     \end{subfigure}
     \hfill
     \begin{subfigure}[b]{0.48\textwidth}
         \centering
         \includegraphics[width=\textwidth]{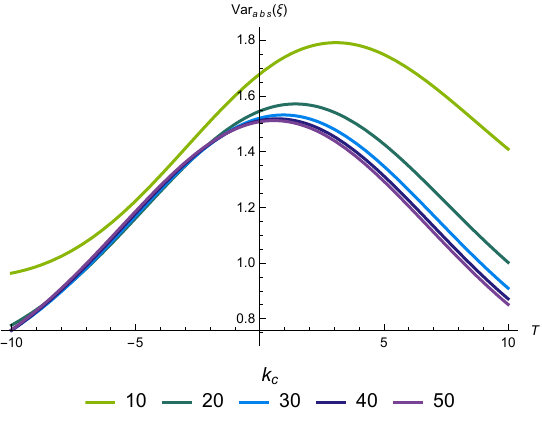}
         \caption{$\text{Var}_{abs}(\xi)$ for $\Lambda_c = 1$, $\sL = 0.1$, $\sk = 1$.}
         \label{fig:varxiabsk}
     \end{subfigure}
     \hfill
     \begin{subfigure}[b]{0.48\textwidth}
         \centering
         \includegraphics[width=\textwidth]{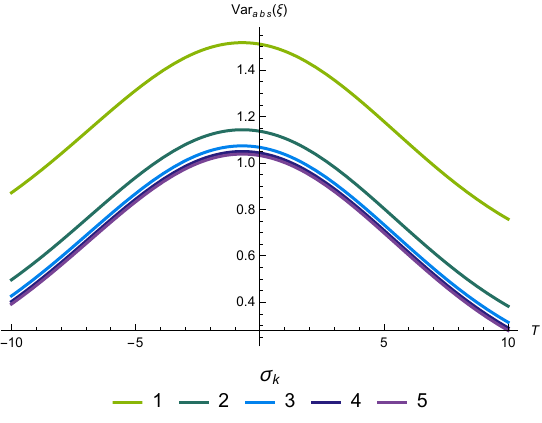}
         \caption{$\text{Var}_{abs}(\xi)$ for $\Lambda_c = 1$, $\sL = 0.1$, $k_c = 40$.}
         \label{fig:varxiabssk}
     \end{subfigure}
        \caption{Plots of $\text{Var}(\xi)$ in the states $\psi_{sc}$ (\cref{fig:varxisck} and \cref{fig:varxiscsk}) and $\psi_{abs}$ (\cref{fig:varxiabsk} and \cref{fig:varxiabssk}). The self-adjoint extension parameter is taken to be $\beta=2$ for $\psi_{sc}$. For $\psi_{abs}$, we choose $\gamma=-2$. In the state $\psi_{sc}$ the variance can be made arbitrarily small everywhere by increasing $\sk$, whereas in the state $\psi_{abs}$ the variance at the origin seems to never go below 1 for any value of $\sk$. As $\langle\hat{\xi}\rangle_{abs}\rightarrow 0$ at the origin the state is not semiclassical close to $T = 0$.}
        \label{fig:varxi}
\end{figure}

\begin{figure}
     \centering
     \begin{subfigure}[b]{0.48\textwidth}
         \centering
         \includegraphics[width=\textwidth]{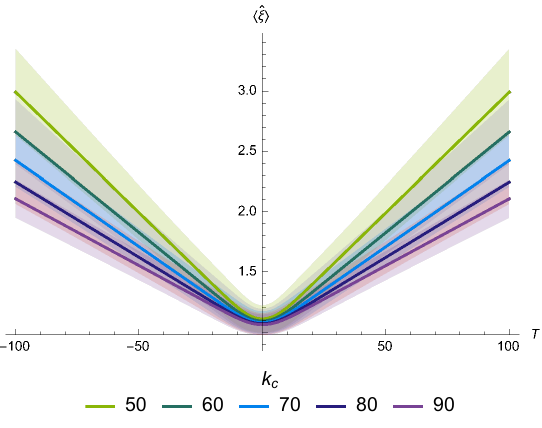}
         \caption{$\langle\hat{\xi}\rangle_{sc}$ for $\Lambda_c = 1$, $\sL = 0.1$, $\sk = 10$, $\beta = 2$ and various $k_c$.}
     \end{subfigure}
     \hfill
     \begin{subfigure}[b]{0.48\textwidth}
         \centering
         \includegraphics[width=\textwidth]{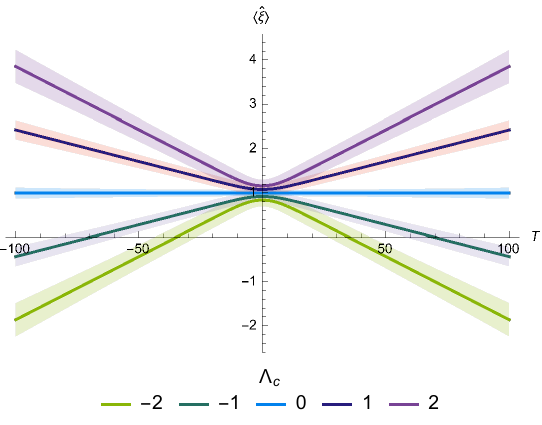}
         \caption{$\langle\hat{\xi}\rangle_{sc}$ for $k_c = 70$, $\sL = 0.1$, $\sk = 10$, $\beta=2$ and various $\Lambda_c$.}
     \end{subfigure}
        \caption{The solid lines show $\langle\hat{\xi}\rangle_{sc}$ and the partially translucent region around a line corresponds to the values within one standard deviation away from the expectation value. One can see that especially close to the origin the behaviour is semiclassical.}
        \label{fig:ev+varsc}
\end{figure}

We will also consider a new state $\psi_{abs}$ defined by
\begin{align}
\label{equ:alphatun}
    \alpha_{abs}(\Lambda,k) = N e^{-i\frac{\gamma}{2}|k|}e^{-\frac{(|k|-k_c)^2}{2\sk^2}}e^{-\frac{(\Lambda-\Lambda_c)^2}{2\sL^2}}\,.
\end{align}
The important difference is that in $\psi_{abs}$ the parameter $\gamma$ can be varied freely without changing the self-adjoint extension, it will always be a state in the theory $\chi(k) = 0$. If this state were also a valid semiclassical state then our statement that in a given quantum theory the sign of the mass of the black/white hole is fixed by the choice of self-adjoint extension would no longer be true. However, by calculating the variance of $\xi$ we can show that this state does not behave semiclassically near $T = 0$. The reason we choose to calculate the variance of $\xi$ instead of $\eta$ is that the latter is the same for $\psi_{sc}$ and $\psi_{abs}$. The calculations are tedious and the resulting expressions can be found in \cref{sec:Avar}. It can be seen from \cref{fig:varxi} that $\psi_{sc}$ exhibits semiclassical behaviour everywhere, as the variance can be tuned down arbitrarily small by increasing $\sk$. This is also illustrated in \cref{fig:ev+varsc}. The same is not true for $\psi_{abs}$ since the variance at $T = 0$ never becomes less than 1, as can also be seen in \cref{fig:varxi}. Thus $\psi_{abs}$ does not behave semiclassically at the origin. 

\begin{figure}
    \centering
    \includegraphics[width=0.5\linewidth]{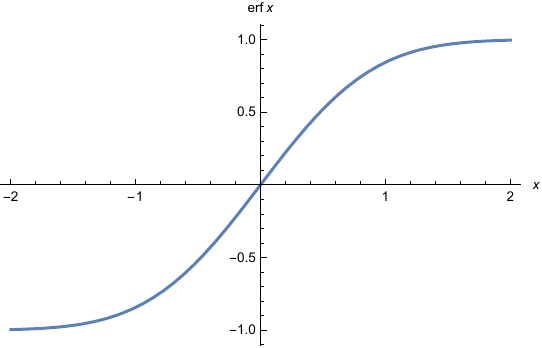}
    \caption{Plot of erf$(x)$.}
    \label{fig:erf}
\end{figure}

\begin{figure}
     \centering
     \begin{subfigure}[b]{0.48\textwidth}
         \centering
         \includegraphics[width=\textwidth]{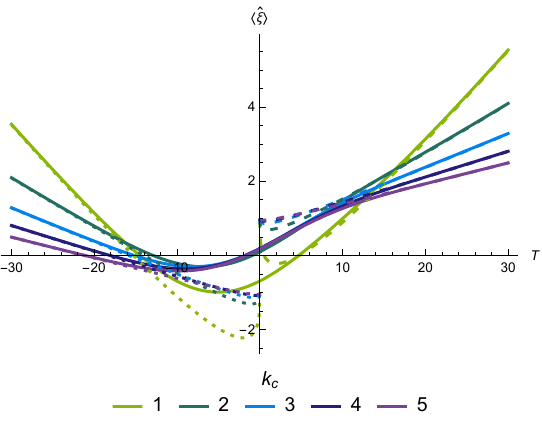}
         \caption{$\langle\hat{\xi}\rangle_{abs}$ with $\Lambda_c = 0.3$, $\sL = 0.1$, $\sk = 0.1$, $\gamma = -2$.}
     \end{subfigure}
     \hfill
     \begin{subfigure}[b]{0.48\textwidth}
         \centering
         \includegraphics[width=\textwidth]{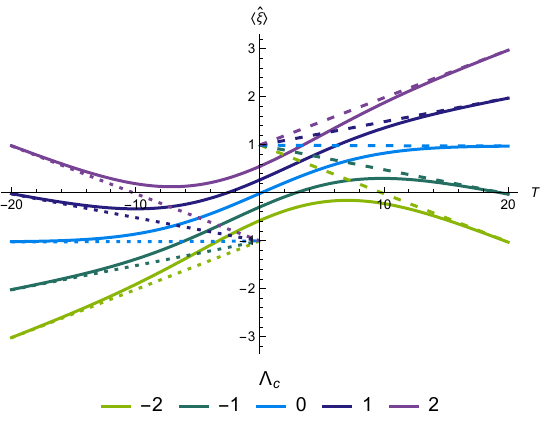}
         \caption{$\langle\hat{\xi}\rangle_{abs}$ with $k_c = 20$, $\sL = 0.1$, $\sk = 0.1$, $\gamma = -2$.}
     \end{subfigure}
        \caption{Plots of $\langle\hat{\xi}\rangle_{abs}$ in solid lines and classical solutions $\xi(T,\Lambda_c,k_c,\xi_0 = 1)$ in  dashed lines and $\xi(T,\Lambda_c,-k_c,\xi_0 = -1)$ in dotted lines. One can see that for asymptotically large positive and negative times the quantum expectation value agrees well with the respective classical solutions. One could interpret the state as describing the process of tunnelling from a semiclassical solution with negative black hole mass to a semiclassical positive mass white hole, where the masses have the same magnitude but opposite signs.}
        \label{fig:evxiabs}
\end{figure}

\begin{figure}
     \centering
     \begin{subfigure}[b]{0.48\textwidth}
         \centering
         \includegraphics[width=\textwidth]{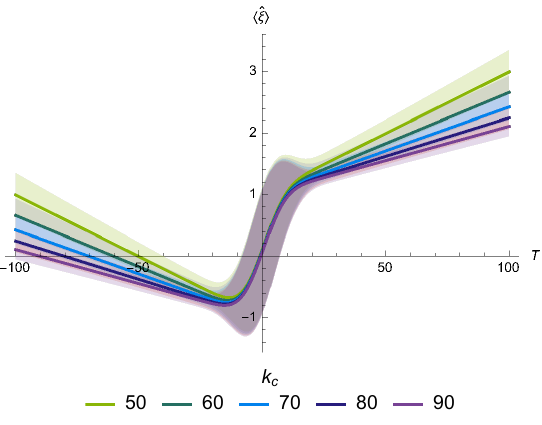}
         \caption{$\langle\hat{\xi}\rangle_{sc}$ for $\Lambda_c = 1$, $\sL = 0.1$, $\sk = 10$, $\gamma = -2$ and various $k_c$.}
     \end{subfigure}
     \hfill
     \begin{subfigure}[b]{0.48\textwidth}
         \centering
         \includegraphics[width=\textwidth]{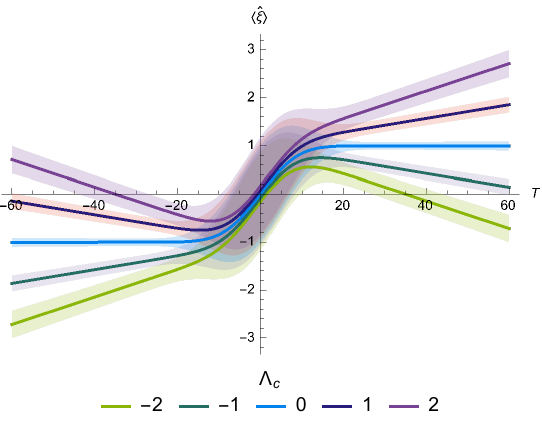}
         \caption{$\langle\hat{\xi}\rangle_{sc}$ for $k_c = 70$, $\sL = 0.1$, $\sk = 10$, $\gamma = -2$ and various $\Lambda_c$.}
     \end{subfigure}
        \caption{The solid lines show $\langle\hat{\xi}\rangle_{abs}$ and the partially translucent region around a line corresponds to the values within one standard deviation away from the expectation value. One can see that there is a region around the origin where the behaviour of the state is highly quantum and the semiclassical interpretation breaks down.}
        \label{fig:ev+varabs}
\end{figure}

(\ref{equ:alphatun}) is not a semiclassical state but corresponds to a tunnelling state that has two asymptotic semiclassical regimes and a highly quantum regime in the middle. It corresponds to the tunnelling of a black hole geometry to a white hole geometry with the sign of the mass switched. To see this, we observe that the calculation of $\langle\hat{\eta}\rangle_{abs}$ is identical to the one of $\langle\hat{\eta}\rangle_{sc}$. The calculation of $\langle\hat{\xi}\rangle_{abs}$ is similar to the calculation of $\langle\hat{\xi}\rangle_{sc}$ up to \cref{equ:absdif}. In the former case this is slightly altered and becomes
\begin{align}
    \langle\hat{\xi}\rangle_{abs} = \int_0^{\infty}\td\eta\int\tDalt\eta^2\,\Bar{\alpha}_1\alpha_2\,e^{\im\lm T}e^{\im\lm\frac{\eta^3}{3k}}\left[\frac{\lp\eta^3}{6k^2}-\frac{\eta}{k^2}+\frac{\gamma}{2}\text{sgn}(k)\right]\,,
\end{align}
i.e., the last term includes $\text{sgn}(k)$. In this expression and in all the calculations in appendix \ref{appendixlabel}, we have defined $\lm = \Lambda_2-\Lambda_1$ and $\lp = \Lambda_2+\Lambda_1$; the integration $\tDalt$ now corresponds to integration over $\lm$, $\lp$ and $k$, and thus differs by a factor of 2 from the previous definition below \cref{equ:dpdef}. In the case of $\psi_{sc}$ this last term resulted in an additive constant $\beta/2$. This is the only difference, as the product $\bar\alpha_1\alpha_2$ is identical to the expression for $\psi_{sc}$ as soon as $k_1 = k_2$ is imposed, so the other terms will give the same result as for $\langle\hat\xi\rangle_{sc}$. The new term gives
\begin{align}
     &  \frac{\gamma}{2}\int_0^{\infty}\td\eta\int\tDalt\eta^2\,\Bar{\alpha}_1\alpha_2\,e^{\im\lm T}e^{\im\lm\frac{\eta^3}{3k}}\text{sgn}(k)\\
    =& \,\frac{\gamma}{4}\intinf\td x\int\tDalt|k|\,\Bar{\alpha}_1\alpha_2\,e^{\im\lm T}e^{\im\lm x}\text{sgn}(x)\\
    =& \,\gamma\frac{N^2}{2}\int\tDalt |k| e^{-\frac{(|k|-k_c)^2}{\sk^2}}e^{-\frac{\lm{}^2}{4\sL^2}}e^{-\frac{(\lp-2\Lambda_c)^2}{4\sL^2}}e^{\im\lm T}\frac{\im}{\lm}\\
    =& -\gamma\frac{N^2}{4}\intinf\frac{\td\lp}{2\pi}\intinf\frac{\td k}{2\pi}|k|\text{erf}(\sL T)e^{-\frac{(|k|-k_c)^2}{\sk^2}}e^{-\frac{(\lp-2\Lambda_c)^2}{4\sL^2}}\\
    =& -\frac{\gamma}{2}\,\text{erf}(\sL T)\,.
\end{align}
The plot of the error function shown in \cref{fig:erf} illustrates the meaning of this alteration to the expectation value of $\hat{\xi}$. For early times $T\ll 0$ the expectation values agree $\langle\hat{\xi}\rangle_{sc}\approx\langle\hat{\xi}\rangle_{abs}$ if $\gamma = \beta$ is chosen. For very late times $T\gg 0$ the new term asymptotes to $- \gamma/2$. In this limit the state again shows semiclassical behaviour as can be seen in the calculation of the variance. The classical solution it corresponds to now has mass $-2\pi\gamma k_c^2$ and is time-reversed. This is visualised in \cref{fig:evxiabs}. The state $\psi_{abs}$, with $\gamma = -2$, could thus be seen as describing the process of tunnelling from a state with negative black hole mass to one with positive white hole mass with a nonsemiclassical regime in the middle. The highly quantum nature of this tunnelling process is illustrated in \cref{fig:ev+varabs}. Of course the same calculation could be done with positive $\gamma$ to give a similar state but with the expectation value converging asymptotically to a classical solution with positive mass for large negative times and to a solution with negative mass for large positive times.

This example demonstrates that the separation of negative and positive mass states into distinct self-adjoint extensions, i.e., different quantum theories, does not persist beyond the semiclassical limit. The full (minisuperspace) theory does contain states connecting negative and positive mass asymptotic solutions. Given that the validity of our minisuperspace approach outside of the semiclassical limit is uncertain, it does not seem clear whether we should expect to see states like $\psi_{abs}$ in full quantum gravity. They would certainly seem troubling since they do not conserve energy as measured by the mass in the classical limit.


\section{Discussion}
\label{sec:dis}
We have shown that imposing unitarity in unimodular time is a mechanism for singularity resolution in quantum Schwarzschild--(Anti-)de Sitter spacetime. Since a self-adjoint extension of the Hamiltonian needs to be chosen, the resulting quantum theory is not unique; using symmetry arguments we can reduce the freedom in this choice to a one-parameter family of quantum theories, each of which correspond to a different boundary condition at the singularity. All these theories resolve all classical singularities, but the sign of the parameter $\beta$ dictates the sign of the mass of semiclassical black/white hole quantum spacetimes. In this sense, at least when restricting to semiclassical Gaussian states we can choose between theories with only positive mass states, a theory with only zero mass states (corresponding to (A)dS with a curious type of ``resolution'' of coordinate singularities), or theories with only negative mass states. The first type of theories seem to be the physically most well-behaved ones, and are hence of main interest in our study. Going away from semiclassicality we have shown that there are tunnelling states mixing positive and negative mass spacetimes. Remarkably, in all these cases we were able to derive analytical expressions for expectation values of metric components, giving a concrete proposal for a quantum-corrected extension of the classical Schwarzschild--(A)dS metric. It would be interesting to investigate detailed properties of this semiclassical quantum spacetime and to compare it with other proposals for quantum-corrected effective black hole geometries.

The non-uniqueness of the boundary condition at the singularity is in contrast to what happens in AdS/CFT, where all information of the theory is encoded in boundary data. A quantity such as the sign of $\beta$, which determines the possible sign of the mass $M$, can be inferred from boundary data. AdS/CFT alone does not impose a restriction on the sign of $M$. In AdS spacetimes black holes with negative mass can in principle exist, however in AdS/CFT they are generally not considered physical because they lead to naked singularities and instability in the dual CFT \cite{HorowitzMyers,Gubser:2000nd}. For a recent discussion of this see \cite{Katoch:2023dfh}. In contrast, our theory with $\beta>0$ inherently includes only positive masses, so that no other criteria have to be imposed. Moreover, while the CFT in AdS/CFT is unitary, there is no direct holographic analogue to imposing unitarity in unimodular time in the bulk theory. Additionally, our model is not limited to negative cosmological constants, whereas AdS/CFT is traditionally formulated in asymptotically AdS spaces. However, there has also been significant work exploring holography in de Sitter space \cite{Strominger:2001pn,Witten:2001kn,Anninos:2017eib,Leuven:2018ejp}.

Unlike other minisuperspace models that impose unitarity with respect to unimodular time $T$ \cite{gielen_black_2024,gielen_singularity_2020,gielen_unitarity_2022-1}, evolution in $T$ is not necessarily timelike in our case. Unitary radial evolution in the context of a Schwarzschild black hole without a cosmological constant has recently been studied in \cite{Magueijo:2025dgy}. At least for $\beta>0$, $T$ remains timelike near the singularity. In general, imposing unitarity of $T$ evolution in our theory can not simply be justified by the requirement of unitary time evolution. If we accept the classical generalisation of the Hamiltonian framework to situations where the evolution is in a spacelike direction, then self-adjointness of $\hat{H}_S$ can be justified from observing that we have the constraint $\hat{H}_S=-\hat\Lambda$ at the quantum level, and $\hat\Lambda$ should correspond to a self-adjoint observable; hence $\hat{H}_S$ should also be represented as a self-adjoint operator. Another classically observable quantity in our model is the black/white hole mass. We have not found an operator whose expectation value corresponds to this observable mass, but it would be interesting to construct such an operator and see if it is self-adjoint in our theory, or what a quantum theory where it is self-adjoint would look like. Other operators that would be interesting to study in the quantum theories constructed  here  are quantum versions of curvature invariants, such as the Kretschmann scalar (\ref{eq:kretschmann}), which might be even better indicators of singularity resolution than the metric coefficients we have focused on here.

In our minisuperspace approach, we have excluded inhomogeneous degrees of freedom prior to quantisation, and solutions behaving highly nonclassically can presumably not be trusted too much. It would be important to include such degrees of freedom into a more complete quantisation that can also take perturbations of the simple spherically symmetric spacetime into account. In most of our work we have restricted ourselves to semiclassical states that let us be agnostic about what the full quantum gravity theory looks like, assuming it leads to unitary evolution in unimodular time $T$, at least in the semiclassical limit. 

The Schwarzschild--(Anti-)de Sitter geometry studied here is the simplest black hole model that includes a cosmological constant. It would be interesting to study more complicated black hole spacetimes, including charge or angular momentum, to see whether imposing unitarity in $T$ is a more general mechanism for singularity resolution in black holes. In particular, it would be interesting to investigate whether the presence of a Cauchy horizon qualitatively alters the quantum behaviour of the spacetime, as recently suggested in \cite{Blacker:2023ezy}. In their analysis, semiclassicality broke down near the singularity—at least for their choice of clock—preventing a reliable probe of the near-singularity regime.

Another possible avenue of future study could be to include Hawking radiation to study the implications of this model on the black hole information loss paradox. The resolution of the singularity and extension of the spacetime by a white hole to the future of the black hole would suggest that information could pass through the quantum region replacing the classical singularity and be retrieved on the other side, although understanding this in detail might still require a full quantum gravity treatment.

\subsection*{Acknowledgements}
We would like to thank Jorma Louko for enlightening discussions on self-adjoint extensions and on the results of \cite{Daughton:1998aa}. The work of SG is funded by the Royal Society
through the University Research Fellowship Renewal
URF$\backslash$R$\backslash$221005.

\appendix

\section{Appendix: Technical details of calculations}
\label{appendixlabel}

\subsection{Expectation values}
\label{sec:Aev}
In this part of the appendix we will give the details of the calculations of the expectation values discussed in the main body of the text. First we will calculate the expectation value of $\hat{\eta}$. For simplicity we will change the $\Lambda$ coordinates to $\lambda^{-} = \Lambda_2-\Lambda_1$ and $\lambda^{+} = \Lambda_2+\Lambda_1$ so that the integration measure changes as
\begin{equation}
\int \frac{\td\Lambda_1\,\td\Lambda_2}{(2\pi)^2} = \frac{1}{2}\int \frac{\td\lambda^+\,\td\lambda^-}{(2\pi)^2}\,. 
\end{equation}
In this appendix we will use the shorthands
\begin{equation}
\int\tD = \int \frac{\td k_1\,\td k_2\,\td\lambda^+\,\td\lambda^-}{(2\pi)^4}\,,\quad  \int\tDalt= \int \frac{\td k\,\td\lambda^+\,\td\lambda^-}{(2\pi)^3}\,.
\end{equation}
We will also write $\alpha_i$ (with $i=1$ or $i=2$) as short for $\alpha_{sc}(\Lambda_i,k_i)$ or $\alpha_{sc}(\Lambda_i,k)$ with $\alpha_{sc}(\Lambda,k)$ defined in \cref{equ:alphasc}.

Using the inner product (\ref{equ:ipxxi}) and the definition of the state (\ref{equ:generalwf}) with (\ref{equ:alphasc}), the expectation value of $\hat{\eta}$ can then be calculated as follows:
\begin{align}
    \langle\hat{\eta}\rangle_{sc} =& \int_0^{\infty}\td\eta\int_{-\infty}^{\infty}\td\xi\int\tD\,\eta^3\,\Bar{\alpha_1}\alpha_2\,e^{\im \lambda^- T}e^{\im \left[(k_2-k_1)\xi+\left(\frac{\Lambda_2}{k_2}-\frac{\Lambda_1}{k_1}\right)\frac{\eta^3}{3}-\left(\frac{1}{k_2}-\frac{1}{k_1}\right)\eta\right]}\\
    =& \int_0^{\infty}\td\eta\int\tDalt\,\eta^3\,\Bar{\alpha_1}\alpha_2\,e^{\im \lambda^-\left( T + \frac{\eta^3}{3k}\right)}\\
    =&\, \frac{1}{2}\int_{-\infty}^{\infty}\td\eta\int\tDalt\,|\eta|^3\Bar{\alpha_1}\alpha_2\,e^{\im \lambda^-\left( T + \frac{\eta^3}{3k}\right)}\\
    =& \,\frac{1}{2}\int_{-\infty}^{\infty}\td y\int\tDalt\,\sqrt[3]{3|k|^4|y|}\,\Bar{\alpha_1}\alpha_2\,e^{\im\lambda^{-}T}\,e^{\im\lambda^{-}y}\\
    =& -\frac{\sqrt[3]{3}N^2}{2}\Gamma\left(\frac{4}{3}\right)\int\tDalt\,\frac{|k|^{4/3}}{|\lm|^{4/3}}e^{-\frac{(|k|-k_c)^2}{\sigma_k^2}}e^{-\frac{\lm{}^2}{4\sigma_{\Lambda}^2}}e^{-\frac{(\lp-2\Lambda_c)^2}{4\sigma_{\Lambda}^2}}e^{\im\lm T}\\
    \begin{split}
    =& \,N^2\frac{\sqrt[3]{3\sL^2\sk^7}\Gamma\left(\frac{2}{3}\right)}{4\pi^2}{}_1F_1\left(-\frac{1}{6},\frac{1}{2},-\sigma_{\Lambda}^2T^2\right)\left[\Gamma\left(\frac{7}{6}\right){}_1F_1\left(-\frac{2}{3},\frac{1}{2},-\frac{k_c^2}{\sigma_k^2}\right)\right.\\
    &\left.+2\Gamma\left(\frac{5}{3}\right)\frac{k_c}{\sigma_k}{}_1F_1\left(-\frac{1}{6},\frac{3}{2},-\frac{k_c^2}{\sigma_k^2}\right)\right]\,.
    \end{split}
\end{align}
Going from the second to the third line in this calculation we have made use of the symmetry $\bar{\alpha}_1(k)\alpha_2(k) = \bar{\alpha}_1(-k)\alpha_2(-k)$. In the next line we have defined $y=\frac{\eta^3}{3k}$. The Fourier transform of $|y|^{\frac{1}{3}}$ is the homogeneous distribution $-\Gamma\left(\frac{4}{3}\right)|\lm|^{-\frac{4}{3}}$, which is defined in a similar way to the Cauchy principal value. For details on homogeneous distributions see \cite{gel'fand1964generalized}. For this calculation the details are not important as in the last step the $\lm$ integral is a Fourier transform of a Gaussian times this homogeneous distribution and can therefore be calculated by performing the Fourier transforms of the two elements separately and then calculating the convolution. The expression ${}_1F_1$ denotes the confluent hypergeometric function.

The expectation value of $\hat{\xi}$ can be calculated in a similar way:
\begin{align}
    \langle\hat\xi\rangle_{sc} =& \int_0^{\infty}\td \eta\int_{-\infty}^{\infty}\td\xi\int\tD\,\eta^2\xi\,\Bar{\alpha}_1\alpha_2\,e^{\im\lm T}e^{\im \left[(k_2-k_1)\xi+\left(\frac{\Lambda_2}{k_2}-\frac{\Lambda_1}{k_1}\right)\frac{\eta^3}{3}-\left(\frac{1}{k_2}-\frac{1}{k_1}\right)\eta\right]}\\
    =& \,2\pi \im\int_0^{\infty}\td\eta\int\tD\,\eta^2\,\Bar{\alpha}_1\alpha_2\,e^{\im\lm T}e^{\im \left[\left(\frac{\Lambda_2}{k_2}-\frac{\Lambda_1}{k_1}\right)\frac{\eta^3}{3}-\left(\frac{1}{k_2}-\frac{1}{k_1}\right)\eta\right]}\delta'(k_1-k_2)\\
    \label{equ:absdif}
    =& \int_0^{\infty}\td\eta\int\tDalt\,\eta^2\,\Bar{\alpha}_1\alpha_2\,e^{\im\lm T}e^{\im\lm\frac{\eta^3}{3k}}\left[\frac{\lp\eta^3}{6k^2}-\frac{\eta}{k^2}+\frac{\beta}{2}\right]\\
    =& \,\frac{1}{2}\int_{-\infty}^{\infty}\td\eta\int\tDalt\,\eta^2\,\Bar{\alpha}_1\alpha_2\,e^{\im\lm T}e^{\im\lm\frac{\eta^3}{3k}}\left[\frac{\lp|\eta|^3}{6k^2}-\frac{|\eta|}{k^2}+\frac{\beta}{2}\right]\\
    =& \,\frac{1}{2}\int_{-\infty}^{\infty}\td y\int\tDalt\,\Bar{\alpha}_1\alpha_2\,e^{\im\lm T}e^{\im\lm y}\left[\frac{\lp}{2}|y|-\frac{\sqrt[3]{3}|y|^{1/3}}{|k|^{2/3}}+|k|\frac{\beta}{2}\right]\\
    =& \,\frac{N^2}{2}\int\tDalt\, e^{-\frac{(|k|-k_c)^2}{\sigma_k^2}-\frac{\lm{}^2+(\lp-2\Lambda_c)^2}{4\sigma_{\Lambda}^2}+\im\lm T}\left[\frac{-\lp}{\lm{}^2}+\frac{\sqrt[3]{3} \Gamma\left(\frac{4}{3}\right)}{|\lm{}^2k|^{2/3}}+|k|\beta \pi\delta(\lm)\right]\\
    \begin{split}
    =&\, \frac{N^2}{2}\int_{-\infty}^{\infty}\frac{\td \lp}{2\pi}\int_{-\infty}^{\infty}\frac{\td k}{2\pi}\,e^{-\frac{(|k|-k_c)^2}{\sigma_k^2}}e^{-\frac{(\lp-2\Lambda_c)^2}{4\sigma_{\Lambda}^2}}\left[\frac{\lp}{2}\left(\frac{1}{\sqrt{\pi}\sigma_{\Lambda}}e^{-\sigma_{\Lambda}^2T^2}+T\text{erf}(\sigma_{\Lambda}T)\right)\right.\\
    &\left.-\frac{\Gamma\left(\frac{2}{3}\right)}{\sqrt{\pi}}\sqrt[3]{\frac{3}{\sL|k|^2}}\,{}_1F_1\left(-\frac{1}{6},\frac{1}{2},-\sigma_{\Lambda}^2T^2\right)+|k|\frac{\beta}{2}\right]
    \end{split}\\
    \begin{split}
    =& \,\frac{N^2}{2\pi}\left[\frac{\Lambda_c}{2}\sigma_k\text{erfc}\left(-\frac{k_c}{\sigma_k}\right)\left(\frac{1}{\sqrt{\pi}}e^{-\sigma_{\Lambda}^2T^2}+\sL T\text{erf}(\sL T)\right)\right.\\
    &\left.-\frac{1}{\pi}\Gamma\left(\frac{2}{3}\right)\sqrt[3]{3\sL^2\sk}\left(3\Gamma\left(\frac{7}{6}\right){}_1F_1\left(\frac{1}{3},\frac{1}{2},-\frac{k_c^2}{\sk^2}\right)\right.\right.\\
    &\left.\left.+\Gamma\left(\frac{2}{3}\right)\frac{k_c}{\sk}{}_1F_1\left(\frac{5}{6},\frac{3}{2},-\frac{k_c^2}{\sk^2}\right)\right){}_1F_1\left(-\frac{1}{6},\frac{1}{2},-\sL^2 T^2\right)\right]+\frac{\beta}{2}\,.
    \end{split}
\end{align}
Here erfc denotes the complementary error function and erf the error function. We see that the self-adjoint extension parameter $\beta$ acts as a constant shift in $\langle\hat\xi\rangle_{sc}$ analogous to the constant shift $\xi_0$ of the classical solution, with the precise identification $\beta=2\xi_0$ expected from classical solutions.

The expectation value $\langle\widehat{\xi/\eta}\rangle_{sc}$ can be calculated as follows:
\begin{align}
    \langle\widehat{\xi/\eta}\rangle_{sc} =& \int_0^{\infty}\td\eta\int_{-\infty}^{\infty}\td\xi\int\tD\, \xi\eta\,\Bar{\alpha}_1\alpha_2\,e^{\im\lm T}e^{\im \left[(k_2-k_1)\xi+\left(\frac{\Lambda_2}{k_2}-\frac{\Lambda_1}{k_1}\right)\frac{\eta^3}{3}-\left(\frac{1}{k_2}-\frac{1}{k_1}\right)\eta\right]}\\
    =& \,\frac{1}{2}\int_{-\infty}^{\infty}\td\eta\int\tDalt\,|\eta|\,\Bar{\alpha}_1\alpha_2\,e^{\im\lm T}e^{\im\lm\frac{\eta^3}{3k}}\left[\frac{\lp|\eta|^3}{6k^2}-\frac{|\eta|}{k^2}+\frac{\beta}{2}\right]\\
    =& \,\frac{1}{2}\int_{-\infty}^{\infty}\td y\int\tDalt\,\Bar{\alpha}_1\alpha_2\,e^{\im\lm T}e^{\im\lm y}\left[\frac{\beta}{2}\frac{|k|^{2/3}}{\sqrt[3]{3|y|}}+\frac{\lp}{2}\frac{|y|^{2/3}}{\sqrt[3]{3|k|}}-\frac{1}{|k|}\right]\\
    \begin{split}
    =&\, \frac{N^2}{2}\int\tDalt\,e^{-\frac{(|k|-k_c)^2}{\sk^2}}e^{-\frac{\lm{}^2}{4\sL^2}}e^{-\frac{(\lp-2\Lambda_c)^2}{4\sL^2}}e^{\im\lm T}\left[\frac{\beta}{2\sqrt[3]{3}}\Gamma\left(\frac{2}{3}\right)\frac{|k|^{2/3}}{|\lm|^{2/3}}\right.\\
    &\left.-\frac{\lp}{3^{5/6}}\Gamma\left(\frac{2}{3}\right)\frac{1}{|k|^{1/3}|\lm|^{5/3}}-\frac{1}{|k|}2\pi\delta(\lm)\right]
    \end{split}\\
    \begin{split}
    =&\, \frac{N^2}{2\sqrt{\pi}}\int_{-\infty}^{\infty}\frac{\td\lp}{2\pi}\int_{-\infty}^{\infty}\frac{\td k}{2\pi}\,e^{-\frac{(|k|-k_c)^2}{\sk^2}}e^{-\frac{(\lp-2\Lambda_c)^2}{4\sL^2}}\left[\frac{\beta}{2}\Gamma\left(\frac{1}{3}\right)|k|^{2/3}\sqrt[3]{\frac{\sL}{3}}\right.\\
    &\left.\times{}_1F_1\left(\frac{1}{6},\frac{1}{2},-\sL^2T^2\right)+\frac{\lp}{2|k|^{1/3}}\frac{\Gamma\left(\frac{5}{6}\right)}{\sqrt[3]{3\sL^2}}{}_1F_1\left(-\frac{1}{3},\frac{1}{2},-\sL^2T^2\right)-\frac{\sqrt{\pi}}{|k|}\right]\,,
    \end{split}\\
    \begin{split}
    =&\, \frac{N^2}{4\pi^2}\left[\beta\sqrt[3]{\frac{\sL^4\sk^5}{3}}{}_1F_1\left(\frac{1}{6},\frac{1}{2},-\sL^2T^2\right)\left(\frac{\sqrt{\pi}}{2^{2/3}}\Gamma\left(\frac{2}{3}\right){}_1F_1\left(-\frac{1}{3},\frac{1}{2},-\frac{k_c^2}{\sk^2}\right)\right.\right.\\
    &\left.\left.+3\frac{k_c}{\sk}\Gamma\left(\frac{4}{3}\right)^2{}_1F_1\left(\frac{1}{6},\frac{3}{2},-\frac{k_c^2}{\sk^2}\right)\right)+\Lambda_c\Gamma\left(\frac{5}{6}\right){}_1F_1\left(-\frac{1}{3},\frac{1}{2},-\sL^2T^2\right)\right.\\
    &\left.\times\sqrt[3]{\frac{\sL\sk^2}{3}}\left(\Gamma\left(\frac{1}{3}\right){}_1F_1\left(\frac{1}{6},\frac{1}{2},-\frac{k_c^2}{\sk^2}\right)+2\frac{k_c}{\sk}\Gamma\left(\frac{5}{6}\right){}_1F_1\left(\frac{2}{3},\frac{3}{2},-\frac{k_c^2}{\sk^2}\right)\right)\right.\\
    &\left.-\sqrt{\pi}\sL\intinf\td k\frac{1}{|k|}e^{-\frac{(|k|-k_c)^2}{\sk^2}}\right]\,.
    \end{split}
\end{align}
The second line of the calculation includes two steps: the $\xi$ integration resulted in $2\pi\im\delta'(k_1-k_2)$, and then we integrated by parts so that either the $k_1$ or $k_2$ integral could be performed. The partial integration results in terms of order $1/k^2$, which is not a distribution. To make it well-defined, we have to again define it as a homogeneous distribution (again for details see \cite{gel'fand1964generalized}). After the substitution $y=\frac{3\eta^3}{k}$ the divergences are in the Fourier transform of the $|y|^{\frac{2}{3}}$ term and in the $1/|k|$ term; these have to be understood as homogeneous distributions. Concretely, the integral in the last line has to be understood as
\begin{align}
\label{equ:homdisabsk}
    \intinf\td k\frac{1}{|k|}e^{-\frac{(|k|-k_c)^2}{\sk^2}} := \intinf\td k\frac{1}{|k|}\left(e^{-\frac{(|k|-k_c)^2}{\sk^2}}-e^{-\frac{k_c^2}{\sk^2}}\right)\,,
\end{align}
so that the logarithmic divergence at $k=0$ has been removed.

\subsection{Variance}
\label{sec:Avar}
 To calculate of the variance of $\hat{\xi}$, we start with the expectation value of $\hat\xi^2$ in the state $\psi_{sc}$:
\begin{align}
    \langle\hat\xi^2\rangle_{sc} =& \int_0^{\infty}\td\eta\intinf\td\xi\int\tD\,\eta^2\xi^2\,\Bar{\alpha}_1\alpha_2\,e^{\im\lm T}e^{\im \left[(k_2-k_1)\xi+\left(\frac{\Lambda_2}{k_2}-\frac{\Lambda_1}{k_1}\right)\frac{\eta^3}{3}-\left(\frac{1}{k_2}-\frac{1}{k_1}\right)\eta\right]}\\
    =& -2\pi\int_0^{\infty}\td\eta\int\tD\,\eta^2\,\delta''(k_2-k_1)\,\Bar{\alpha}_1\alpha_2\,e^{\im\lm T}e^{\im \left[\left(\frac{\Lambda_2}{k_2}-\frac{\Lambda_1}{k_1}\right)\frac{\eta^3}{3}-\left(\frac{1}{k_2}-\frac{1}{k_1}\right)\eta\right]}\\
    \begin{split}
    =& \int_0^{\infty}\td\eta\int\tDalt\,\eta^2\Bar{\alpha}_1\alpha_2e^{\im\lm T}e^{\im\lm \frac{\eta^3}{3k}}\left[\frac{1}{k^4}\left(\eta-\frac{\lp\eta^3}{6}\right)^2-\im\frac{\lm\eta^3}{6k^3}+\frac{1}{2\sk^2}\right.\\
    &\left.-\frac{k_c}{\sk^2}\delta(k)+\beta\left(\frac{\lp\eta^3}{6k^2}-\frac{\eta}{k^2}\right)+\frac{\beta^2}{4}\right]
    \end{split}\\
    \begin{split}
    =& \,\frac{1}{2}\intinf\td y\int\tDalt\,\Bar{\alpha}_1\alpha_2\,e^{\im\lm T}e^{\im\lm y}\left[\frac{\left(\sqrt[3]{3}|y|^{1/3}-\frac{\lp}{2}|k|^{2/3}|y|\right)^2}{|k|^{7/3}}-\im\frac{\lm}{2|k|}y\right.\\
    &\left.+\frac{|k|}{2\sk^2}-\frac{k_c}{\sk^2}|k|\delta(k)+\beta\left(\frac{\lp}{2}|y|-\frac{\sqrt[3]{3}|y|^{1/3}}{|k|^{2/3}}\right)+\frac{\beta^2}{4}|k|\right]
    \end{split}\\ 
    \begin{split}
    =& \,\frac{N^2}{2}\int\tDalt\,e^{-\frac{(|k|-k_c)^2}{\sk^2}}e^{-\frac{\lm{}^2}{4\sL^2}}e^{-\frac{(\lp-2\Lambda_c)^2}{4\sL^2}}e^{\im\lm T}\left[-\frac{3^{7/6}}{|k|^{7/3}|\lm|^{5/3}}\Gamma\left(\frac{5}{3}\right)\right.\\
    &\left.+\lp\frac{3^{5/6}}{|k|^{5/3}|\lm|^{7/3}}\Gamma\left(\frac{7}{3}\right)+\beta\left(-\frac{\lp}{|\lm|^2}+\frac{\sqrt[3]{3}}{|k|^{2/3}|\lm|^{4/3}}\Gamma\left(\frac{4}{3}\right)\right)\right.\\
    &\left.-\frac{(\lp)^2\pi}{2|k|}\delta''(\lm)-\frac{\lm \pi}{|k|}\delta'(\lm)+\left(\frac{|k|}{\sk^2}-\frac{2k_c}{\sk^2}|k|\delta(k)+\frac{\beta^2}{2}|k|\right)\pi\delta(\lm)\right]
    \end{split}\\ 
    \begin{split}
    =& \,\frac{N^2}{2\sqrt{\pi}}\intinf\frac{\td k}{2\pi}\intinf\frac{\td \lp}{2\pi}\,e^{-\frac{(|k|-k_c)^2}{\sk^2}}e^{-\frac{(\lp-2\Lambda_c)^2}{4\sL^2}}\left[\sqrt{\pi}|k|\left(\frac{1}{2\sk^2}+\frac{\beta^2}{4}\right)\right.\\
    &\left.+\frac{3^{2/3}}{|k|^{7/3}\sL^{2/3}}\Gamma\left(\frac{5}{6}\right){}_1F_1\left(-\frac{1}{3},\frac{1}{2},-\sL^2T^2\right)+\frac{\lp{}^2\sqrt{\pi}}{4|k|}\left(T^2+\frac{1}{2\sL^2}\right)+\frac{\sqrt{\pi}}{2|k|}\right.\\
    &\left.-\frac{3^{1/3}\Gamma\left(\frac{2}{3}\right)}{|k|^{2/3}\sL^{1/3}}\,\beta\,{}_1F_1\left(-\frac{1}{6},\frac{1}{2},-\sL^2T^2\right)-\frac{3^{1/3}\Gamma\left(\frac{7}{6}\right)}{|k|^{5/3}\sL^{4/3}}\,\lp\,{}_1F_1\left(-\frac{2}{3},\frac{1}{2},-\sL^2T^2\right)\right.\\
    &\left.+\frac{\beta\lp}{2\sL}\left(e^{-\sL^2T^2}+\sqrt{\pi}\sL T \text{erf}(\sL T)\right)\right]
    \end{split}\\ 
    \begin{split}
    =& \frac{N^2\sL}{4\pi^2}\left[K_{\frac{7}{3}}\frac{3^{2/3}}{\sL^{2/3}}{}_1F_1\left(-\frac{1}{3},\frac{1}{2},-\sL^2T^2\right)-K_{\frac{5}{3}}\sqrt[3]{\frac{3}{\sL^4}}\Lambda_c{}_1F_1\left(-\frac{2}{3},\frac{1}{2},-\sL^2T^2\right)\right.\\
    &\left.-K_{\frac{2}{3}}\frac{\beta}{2}\sqrt[3]{\frac{3}{\sL}}\,{}_1F_1\left(-\frac{1}{6},\frac{1}{2},-\sL^2T^2\right)+K_0\frac{\beta\Lambda_c}{\sL}\left(e^{-\sL^2T^2}+\sqrt{\pi}\sL T \text{erf}(\sL T)\right)\right.\\
    &\left.+K_{-1}\sqrt{\pi}\left(\frac{1}{2\sk^2}+\frac{\beta^2}{4}\right)+K_{1}\sqrt{\pi}\left(\left(\Lambda_c^2+\frac{\sL^2}{2}\right)\left(T^2+\frac{1}{2\sL^2}\right)+\frac{1}{2}\right)\right]\,.
    \end{split}
\end{align}
In the last expression we have defined the following functions of $k_c$ and $\sk$:
\begin{align}
    K_{\frac{7}{3}} &:= \Gamma\left(\frac{5}{6}\right)\sk^{-\frac{4}{3}}\left[\Gamma\left(-\frac{2}{3}\right){}_1F_1\left(\frac{7}{6},\frac{1}{2},-\frac{k_c^2}{\sk^2}\right)+2\frac{k_c}{\sk}\Gamma\left(-\frac{1}{6}\right){}_1F_1\left(\frac{5}{3},\frac{3}{2},-\frac{k_c^2}{\sk^2}\right)\right]\,,\\
    K_{\frac{5}{3}} &:= 2\Gamma\left(\frac{7}{6}\right)\sk^{-\frac{2}{3}}\left[\Gamma\left(-\frac{1}{3}\right){}_1F_1\left(\frac{5}{6},\frac{1}{2},-\frac{k_c^2}{\sk^2}\right)+2\frac{k_c}{\sk}\Gamma\left(\frac{1}{6}\right){}_1F_1\left(\frac{4}{3},\frac{3}{2},-\frac{k_c^2}{\sk^2}\right)\right]\,,\\
    K_{\frac{2}{3}} &:= 4\Gamma\left(\frac{2}{3}\right)\sk^{\frac{1}{3}}\left[3\Gamma\left(\frac{7}{6}\right){}_1F_1\left(\frac{1}{3},\frac{1}{2},-\frac{k_c^2}{\sk^2}\right)+\frac{k_c}{\sk}\Gamma\left(\frac{2}{3}\right){}_1F_1\left(\frac{5}{6},\frac{3}{2},-\frac{k_c^2}{\sk^2}\right)\right]\,,\\
    K_{-1} &:= \sk^2 e^{-\frac{k_c^2}{\sk^2}}+\sqrt{\pi}\,k_c\sk\,\left(1+\text{erf}\left(\frac{k_c}{\sk}\right)\right)\,,\\
    K_0 &:= \sqrt{\pi}\sk\left(1+\text{erf}\left(\frac{k_c}{\sk}\right)\right)\,,\qquad
    K_{1} := \intinf\td k \frac{1}{|k|}e^{-\frac{(|k|-k_c)^2}{\sk^2}}\,.
\end{align}
$K_{1}$ is defined as in \cref{equ:homdisabsk}. 

The calculation of the variance of $\xi$ in the state $\psi_{abs}$ is very similar: we obtain
\begin{align}
    \begin{split}
    \langle\hat\xi^2\rangle_{abs}  =&\int_0^{\infty}\td\eta\int\tDalt\,\eta^2\Bar{\alpha}_1\alpha_2\,e^{\im\lm T}e^{\im\lm \frac{\eta^3}{3k}}\left[\frac{1}{k^4}\left(\eta-\frac{\lp\eta^3}{6}\right)^2-\im\frac{\lm\eta^3}{6k^3}+\frac{1}{2\sk^2}\right.\\
    &\left.-\frac{k_c}{\sk^2}\delta(k)+\gamma\,{\rm sgn}(k)\left(\frac{\lp\eta^3}{6k^2}-\frac{\eta}{k^2}\right)+\frac{\gamma^2}{4}\right]
    \end{split}
\end{align}
and we see that the only difference is in the second to last term, i.e., the one linear in $\gamma$. We will therefore now only calculate this term, as the rest is identical to the calculation above (with $\beta\rightarrow\gamma$ in the last term). We have
\begin{align}
    &  \gamma\int_0^{\infty}\td\eta\int\tDalt\,\eta^2\,\Bar{\alpha}_1\alpha_2\,e^{\im\lm T}e^{\im\lm \frac{\eta^3}{3k}}\text{sgn}(k)\left(\frac{\lp\eta^3}{6k^2}-\frac{\eta}{k^2}\right)\\
    =&\,N^2\frac{\gamma}{2} \intinf\td y\int\tDalt\,e^{-\frac{(|k|-k_c)^2}{\sk^2}}e^{-\frac{\lm{}^2}{4\sL^2}}e^{-\frac{(\lp-2\Lambda_c)^2}{4\sL^2}}e^{\im\lm T}e^{\im\lm y}\left(\frac{\lp}{2}y-\sqrt[3]{\frac{3y}{k^2}}\right)\\
    \begin{split}
    =&\,N^2\frac{\gamma}{2} \intinf\td y\intinf\frac{\td k}{2\pi}\intinf\frac{\td\lp}{2\pi}\,e^{-\frac{(|k|-k_c)^2}{\sk^2}}e^{-\frac{(\lp-2\Lambda_c)^2}{4\sL^2}}e^{-\sL^2(y+T)^2}\frac{\sL}{\sqrt{\pi}}\left(\frac{\lp}{2}y-\sqrt[3]{\frac{3y}{k^2}}\right)
    \end{split}\\
    \begin{split}
    =& \,N^2\frac{\gamma}{2}\intinf\frac{\td k}{2\pi}\intinf\frac{\td\lp}{2\pi}\,e^{-\frac{(|k|-k_c)^2}{\sk^2}-\frac{(\lp-2\Lambda_c)^2}{4\sL^2}}\left(\frac{\sL^{2/3}\Gamma\left(\frac{1}{6}\right)}{3^{2/3}\sqrt{\pi}|k|^{2/3}}{}_1F_1\left(\frac{1}{3},\frac{3}{2},-\sL^2T^2\right)-\frac{\lp}{2}\right)T
    \end{split}\\
    =&\, \frac{N^2\sL}{4\pi^2}\gamma\left(K_{\frac{2}{3}}\frac{\Gamma\left(\frac{7}{6}\right)}{\Gamma\left(\frac{2}{3}\right)}\sqrt[3]{3\sL^2}{}_1F_1\left(\frac{1}{3},\frac{3}{2},-\sL^2T^2\right)-\sqrt{\pi}K_0\Lambda_c \right)T\,.
\end{align}
The functions $K_0$ and $K_{\frac{2}{3}}$ are as defined above. Now that we have calculated the expectation value of $\hat{\xi}^2$ in both states we can study the variance. It can be seen from \cref{fig:varxi} that $\psi_{sc}$ exhibits semiclassical behaviour everywhere, as the variance can be tuned down arbitrarily small by increasing $\sk$. The same is not true for $\psi_{abs}$ since the variance at $T = 0$ never becomes less than 1. Thus $\psi_{abs}$ does not behave semiclassically at the origin.

\bibliographystyle{JHEP.bst}
\bibliography{uniSdS}

\end{document}